\documentclass[numbers,compress,3p,times,review]{elsarticle} 
\usepackage[utf8]{inputenc}
\usepackage{natbib}
\setcitestyle{square, comma, numbers,sort&compress, super}
\usepackage{natbib}
\usepackage{lmodern}
\usepackage[dvipsnames]{xcolor}
\usepackage{booktabs}
\usepackage{multirow}
\usepackage{graphicx}
\usepackage{subcaption}
\usepackage{textcomp, gensymb}
\usepackage{gensymb}
\usepackage{siunitx}
\usepackage{amsmath,amsfonts,amssymb}
\usepackage{epsfig}
\usepackage{cleveref}

\title{In-situ Controller Autotuning by Bayesian Optimization for Closed-loop Feedback Control of Laser Powder Bed Fusion Process}
\journal{Journal of Additive Manufacturing}

\author[inst1]{Bar{\i}\c{s} Kavas\corref{cor1}}
\ead{bkavas@ethz.ch}
\author[inst2,inst3]{Efe C. Balta}
\ead{efe.balta@inspire.ch}
\author[inst1]{Michael R. Tucker}
\ead{mtucker@ethz.ch}
\author[inst2]{Raamadaas Krishnadas}
\ead{raamadaas.krishnadas@inspire.ch}
\author[inst4]{Alisa Rupenyan}
\ead{rupn@zhaw.ch}
\author[inst3]{John Lygeros}
\ead{jlygeros@ethz.ch}
\author[inst1]{Markus Bambach}
\ead{mbambach@ethz.ch}

\cortext[cor1]{Corresponding author}

\address[inst1]{Department of Mechanical and Process Engineering,
ETH Z\"urich, Z\"urich 8092, Switzerland}
\address[inst2]{Control and Automation Group, inspire AG, Zürich 8005, Switzerland}
\address[inst3]{Department of Information Technology and Electrical Engineering, Automatic Control Laboratory, ETH Zürich, Zürich 8092, Switzerland}
\address[inst4]{ZHAW Centre for Artificial Intelligence, Zürich University of Applied Sciences, Switzerland}

\newdefinition{rmk}{Remark}

\begin{document}



\begin{abstract}

Open-loop control of laser powder bed fusion (LPBF) additive manufacturing (AM) has enabled the industrial production of complex and high-criticality parts for aerospace, power generation, medical, transportation, and other industries. This approach relies on static parameter sets obtained through extensive experimentation and a priori simulation on analog parts, with the hope that they remain stable and defect-free once transferred to the production parts.
Closed-loop control of LPBF has the potential to enhance process stability further and reduce defect formation in the face of complex thermal histories, stochastic process noise, hardware drift, and unexpected perturbations. The controllers can be classified based on the spatial and temporal scales in which they operate, designated as layer-to-layer and in-layer controllers.
However, the performance and effectiveness of controllers largely depend on the tuning of their parameters. 
Traditionally, controller tuning has been a manual, expertise-driven process that does not guarantee optimal controller performance and is often constrained by the non-transferability of settings between different systems.
This study proposes the use of Bayesian Optimization (BO), a sample-efficient algorithm, to automate the tuning of an in-layer controller by leveraging the layer-to-layer repetitive nature of the LPBF process.
Two alternative approaches are introduced: online tuning, which adjusts parameters iteratively during the process, and offline tuning, conducted in a representative setup such as laser exposures on a bare metal plate.
The proposed methods are experimentally implemented on an in-layer PI controller and the performance of the resulting tuned controllers is investigated on two different wedge geometries that are prone to overheating. 
The results demonstrate that BO effectively tunes controllers using either method, where both significantly reduced overheating in controlled wedge specimens compared to those uncontrolled.
Notably, this study provides the first printed parts controlled by an in-layer controller and subjected to microstructural analysis in the literature. 
Microstructural findings show the partial presence of lack-of-fusion type porosities induced by the controller assigning insufficient laser power to compensate for the overheating which highlights one of the most significant challenges for the utilization of laser power controllers.
In summary, BO presents a promising method for the automatic tuning of in-layer controllers in LPBF, enhancing control precision and mitigating overheating in production parts. 
Looking forward, BO could extend to broader LPBF settings and related additive manufacturing modalities, potentially transforming controller tuning into a more adaptive and robust process across different machines and materials.

\end{abstract}

\maketitle

\section{Introduction}

Laser Powder Bed Fusion (LPBF) is a prominent additive manufacturing process with applications spanning several industries. 
Despite its widespread adoption, the LPBF process is susceptible to disturbances stemming from its complex multi-physics nature, making it difficult to model the exact behavior of the process. 
There has been an increasing interest in closed-loop feedback control applications on the LPBF process to improve process robustness.
Closed-loop control is proposed to be applied mainly in 2 scales due to the nature of the process: Layer-to-layer and in-layer.
Implementing in-layer closed-loop feedback control systems particularly contains Proportional-Integral-Derivative (PID) controllers due to the high-frequency requirement of the in-layer control objective. 
PID controllers, well-established in various engineering fields, are common in research and practice due to their straightforward design and proven effectiveness.

\subsection{In-layer Controller Applications in LPBF}

The initial adoption of closed-loop feedback in LPBF can be traced back to the seminal paper by Benda et al. \cite{benda1994temperature}, who applied a simple form of control to the LPBF process on iron powder.
Their preliminary experimental study showed the successful feasibility of utilizing an on-axis optical sensor for the dynamic control of the melt pool.
Subsequent efforts use PID controllers~\cite{kruth2007line,kruth2007feedback}. 
These early studies have been limited to qualitative comparisons of basic geometries where they showed the potential of photodiode signal stabilization by actuating the laser power.
The study of Craeghs et al.~\cite{craeghs2010feedback} showed two use cases for the in-layer PI controller.
One of them was an artificial case with smaller hatch spacing to simulate an excessive energy input while the other was the melt pool emission variations in overhanging surfaces. 
They showed improved stability in the melt pool signals in both cases.
Their controller design only included PI terms due to the noise of the photodiode signal which rendered the use of D-term impractical.
Renken et al. implemented an in-layer controller with only the proportional (P) term to control the melt pool. They showed improved thermal stability on a machined bridge structure, where the melt pool undergoes a heat build-up on the thinner cross-section in the uncontrolled case~\cite{renken2018model}. A follow-up study reports the implications of the P-controller application on various vector sizes and a single layer exposure on powder~\cite{Renken2019}.
The controller improves the stability of the melt pool in all cases compared to the predefined feed-forward open-loop laser power inputs.
Syed et al. proposed a controller design by also including the I and D terms presented with a simulation-based study that the PID controller successfully stabilizes in vector variation of the meltpool size in~\cite{hussain2021feedback}.
Shkoruta et al. implemented a PI controller by using the high-speed camera signal and applied the controller both on a single track and a multi-track single-layer where they showed improved stability of the melt pool size compared to the non-controlled case~\cite{shkoruta2022real}.
Finally, Rongxuan Wang et al.~\cite{Wang2023} present a PD control application and experimentally show the stabilization capability of the controller in a single-layer case of vector-to-vector heat accumulation.
Each study referenced to this point follows a common methodological approach toward parameter optimization of PID controllers. 
Specifically, tuning of the gain parameters is conducted completely manually, utilizing heuristic techniques based on an empirical understanding of the effect of individual parameters.

\subsection{The Tuning Challenge of PID Controllers}

The effectiveness of PID controllers is contingent on the tuning and selection of the gains applied to the proportional (P), integral (I), and derivative (D) error terms.
Traditionally, parameter tuning has been a time-consuming heuristic process that is heavily reliant on expertise. 
Moreover, the lack of a universal parameter set that is transferable between different LPBF processes or even varying conditions within the same process exacerbates the tuning challenge.
Varying conditions dictate that the tuning procedure be performed in a setting where the system behavior is representative of the actual printing conditions.
Therefore, the term \textit{offline tuning} refers to the tuning procedure performed under artificial or isolated conditions that are assumed to be sufficiently representative of the actual process conditions, while \textit{online tuning} refers to tuning during the natural process conditions~\cite{park2021online}.
Online and offline tuning approaches each have specific trade-offs for tuning a controller in the manufacturing setting.


While offline tuning is the usual choice for tuning controllers when large changes in the environment or the corresponding system are not expected, the resulting performance is fixed by the selected parameters. ~\cite{memon2020optimal}.
Online, or adaptive tuning is performed close to actual process conditions and can increase performance under changing conditions. However, it may introduce increased variability during the tuning phase.
Besides the challenge arising from the tuning setting, the difference in the order of magnitude of the controller parameters also makes the tuning process lengthy and costly~\cite{campos2009challenges}.
Well-established heuristic methods such as Ziegler-Nichols~\cite{meshram2012tuning} are shown to achieve an effective controller tuning requiring open loop access, without guarantees on the optimality for most cases.
Iterative methods requiring open-loop access are not desired in manufacturing systems due to the safety blocks and time consumption.
Joseph et al. recently published a comprehensive review of PID controller tuning algorithms that summarizes the existing methods for various applications~\cite{joseph2022metaheuristic}.
Among the described algorithms, Bayesian Optimization has been gaining significant traction in various industrial applications.

\subsubsection{Bayesian Optimization for controller tunning}

Bayesian Optimization (BO) is an efficient data-driven optimization algorithm that excels in settings characterized by limited data availability and complex system dynamics.
It operates by modeling the process inputs and outputs as Gaussian processes and selecting sample points that minimize overall uncertainty while promoting exploration.
Since BO uses Gaussian processes, knowledge about the form of the function to model the process is generally not required.
Among the various types of controllers, BO-based autotuning of PID controllers recently gained significant attention~\cite{rupenyan2021performance, khosravi2022safety}.


For example, in motor control, Chen et al. utilized BO to tune the PID controller of an adjustable payload servo motor~\cite{chen2019controller}.
Similarly, Hajieghrary et al.~\cite{Hajieghrary2022} and Fujimoto et al.~\cite{Fujimoto2023} reported improved controller performance with BO-tuned PID controllers in mobile robotic manipulators and servo motor control applications, comparing favorably to traditional tuning methods.
König et al. used a modified version of the BO algorithm to adaptively optimize a cascade PI controller for a rotational axis drive, demonstrating promising outcomes~\cite{konig2021safe} for different operating contexts.
In the following study by Zagorowska et al, the BO algorithm is used to autotune the PID that controls the position of a high-precision motion system and increase the tracking performance.~\cite{zagorowska2023efficient}.
Khosravi et al. showed a 20\% increased controller performance by BO-tuning of the linear axis drive of a CNC grinding machine compared to nominal settings~\cite{khosravi2022safety}.
All the cited work shows the superior performance of the BO autotuned PID controllers in non-linear dynamic systems without requiring a physical model of the system.


The demonstrated effectiveness of the BO algorithm in tuning PID controllers across a variety of complex applications promises to be effective for the distinct challenges for controller tuning of the LPBF process. 
Accurately modeling pyrometer response in the LPBF process is challenging due to complex melt pool dynamics and data acquisition noise, rendering model-based PID tuning impractical. 
Manual tuning of PID controllers is also a labor-intensive process, requiring significant expertise and time without guaranteeing optimal performance. 
Consequently, the use of Bayesian Optimization (BO) as a model-free, sample-efficient algorithm for PID tuning in LPBF offers considerable promise for achieving efficient and high-performance controller tuning.

\subsection{Research Contribution}

So far, to our knowledge, no algorithmic approaches have been reported for PID tuning for within-layer control of the LPBF process.
Given the non-transferable nature of the controllers across different machines and materials, this gap represents a bottleneck for the further utilization of the in-layer controllers. 
Reported controller studies only performed offline tuning procedures where the tuning is manually performed in a representative setting and then applied in the actual process.
There are no reported studies that propose an online controller tuning method without disturbing the LPBF printing process.
In this work, sample efficient controller autotuning procedures for LPBF are presented, based on Bayesian optimization methods. 
Online and offline approaches are experimentally tested on a geometry prone to overheating to illustrate the empirical comparison of both methods and provide a baseline for future research in the field.

The main contributions of this work are:
\begin{itemize}
    \item An autonomous controller tuning procedure that does not require a prestudy of controller parameters and experimental demonstration for in-layer PI controller tuning in the LPBF process,
    \item An online controller tuning procedure that leverages the layer-to-layer nature of the LPBF process with a detailed experimental comparison of printed 3D wedge geometries between the online and offline tuned controller performances.
\end{itemize}

\Cref{sec:material} describes the experimental procedure with detailed descriptions of the characterization methods, hardware, and software used for the reproducibility of the procedures described. 
The proposed autotuning framework and implementation of the BO algorithm are given in \Cref{sec:theory}. 
In \Cref{sec:results}, results are shared where they are discussed in detail in \Cref{sec:discussion}.
Concluding remarks and future directions are given in \Cref{sec:conclusion}.
\section{Materials and methods}
\label{sec:material}

\subsection{Materials}

\subsubsection{LPBF Processing}
An Aconity3D Midi+ (Aconity3D GmbH, Herzogenrath, Germany) LPBF machine \cite{aconity3d} was used in the experiments.
The processing laser was a continuous-wave Gaussian-mode fiber laser with a wavelength of 1080nm and a maximum output of 500W (nLIGHT Alta, Vancouver WA, USA). 
The laser was focused to a beam diameter of $\SI{80}{\micro\meter}$ and a $\SI{30}{\micro\meter}$ layer thickness was used for printing. 
The nominal laser power and scan speed parameters for the given layer thickness and beam diameter for fully-dense microstructure are 150W and 800mm/s, respectively.
The powder used in this study was gas-atomized stainless steel 316L (1.4404) with a particle size distribution of $15-\SI{45}{\micro\meter}$ (CT POWDERRANGE 316LF, Carpenter Additive, Cheshire UK).
The metal plate used to expose the laser for the offline tuning setting was made of S304 steel.

\subsubsection{In-layer Control Hardware and Software}
The PI-based in-layer controller was implemented using the AconityCONTROL hardware and software upgrade package \cite{aconity3d}.

\begin{figure}[ht!]
\begin{center}
\includegraphics[width=0.7\columnwidth]{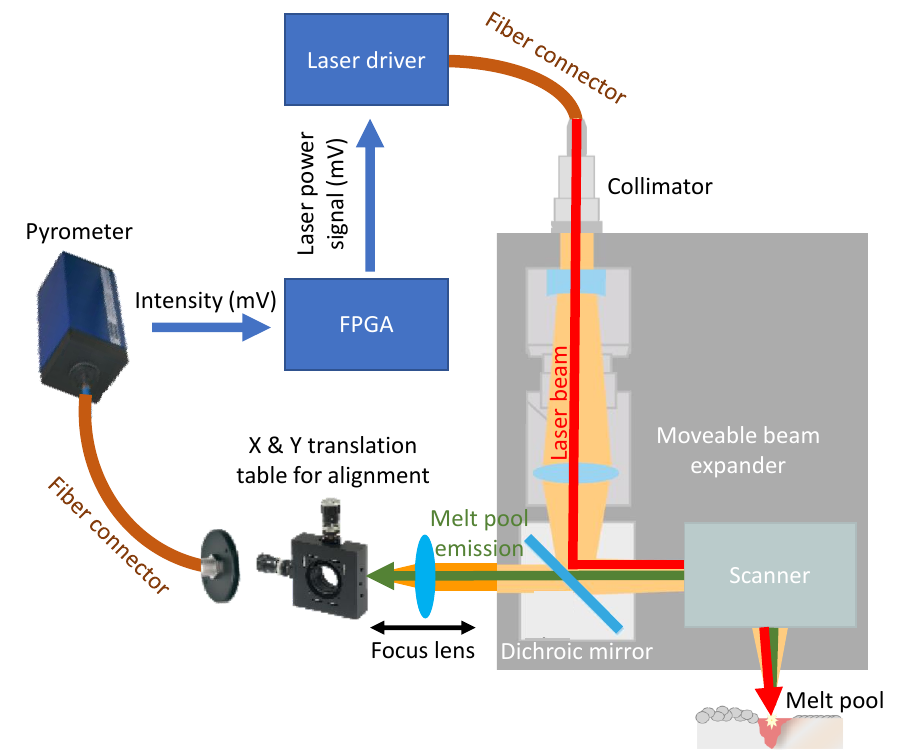}
\caption{Implemented control loop in Aconity Midi+}
\label{fig:pyrometer_setup}
\end{center}
\end{figure}

Schematic of the controller implementation to the optical axis of the machine is given in \Cref{fig:pyrometer_setup}.

The laser beam generated is transmitted to the optical unit via a fiber optic cable. To ensure parallel alignment, the unfocused beam within the cable is passed through a collimator.
The beam is then focused through the beam expander and reflected by a 45-degree dichroic mirror into a galvanometer scanner, which steers the beam across the powder bed. 
The movable beam expander ensures consistent focus regardless of the projection location on the build platform.
Upon sufficient heating, a melt pool is formed.
A portion of the emitted radiation from the vicinity of the melt pool traces back up the same optical path until the dichroic mirror, where it is transmitted onto a pyrometer module. 
The radiation is manually focused with a movable lens and aligned using an X-Y micrometer table. 
The pyrometer converts the intensity of the incipient radiation to an analog voltage at a sampling rate of 100kHz.
The pyrometer used in the setup is Kleiber KG740~\cite{kleiberinfrared} with a wavelength range of 1500 to $\SI{1700}{\nano\meter}$.
The intensity reading is passed to a field-programmable gate array (FPGA) for the embedded feedback signal calculation of the implemented PI controller.
Finally, the generated input signal is fed back into the laser driver during the laser power assignment during the continuing exposure.
The sampling rate matches the clock frequency of the controller PC, therefore the total delay time of the entire loop matches the time length of a single sample, which is $\SI{10}{\micro\second}$.

\subsection{Method}

\begin{figure}[ht!]
\begin{center}
\includegraphics[width=0.6\columnwidth]{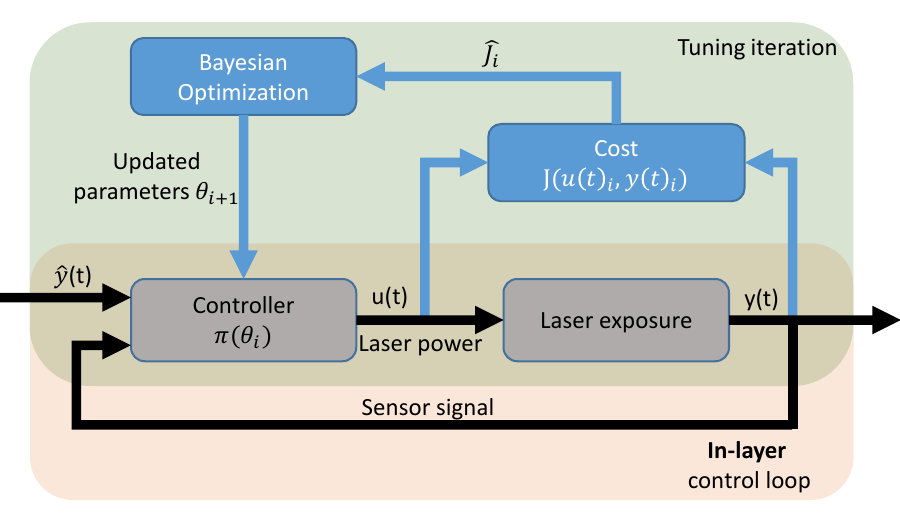}
\caption{Proposed comprehensive schematic for online and offline controller optimization}
\label{fig:optimization_loop}
\end{center}
\end{figure}

The generalized schematic of the proposed auto-tuning procedure is shown in \Cref{fig:optimization_loop}, and is comprised of two main loops.

The lower loop, highlighted in orange, represents the in-layer controller.
The controller actuates the laser power $u(t)$ within a scan vector to match the sensor reading $y(t)$ to the defined reference value using the PI controller and pyrometer feedback.

The upper loop, highlighted in green, represents the auto-tuning iteration loop.
The auto-tuning procedure is initialized with a predefined controller parameter set.
To find updated values, the in-layer control loop is used to expose a single vector with pyrometer feedback.
After exposure, a cost value $\hat{J}_i$ is calculated by using the $u(t)$ and $y(t)$ as inputs.
Then the cost value is used by the BO algorithm to calculate the new set of controller parameters $\theta_{i+1}$ to be passed to the controller.
The tuning iteration loop is repeated either until the cost value is below a target threshold or until a specified number of iterations is reached, after which the parameters are held constant.
The design of the cost function and the BO algorithm used in this study are elaborated in detail in \Cref{sec:theory}.

\subsubsection{Offline and Online Controller Tuning Methods}
\begin{figure}[ht!]
\begin{center}
\includegraphics[width=0.8\columnwidth]{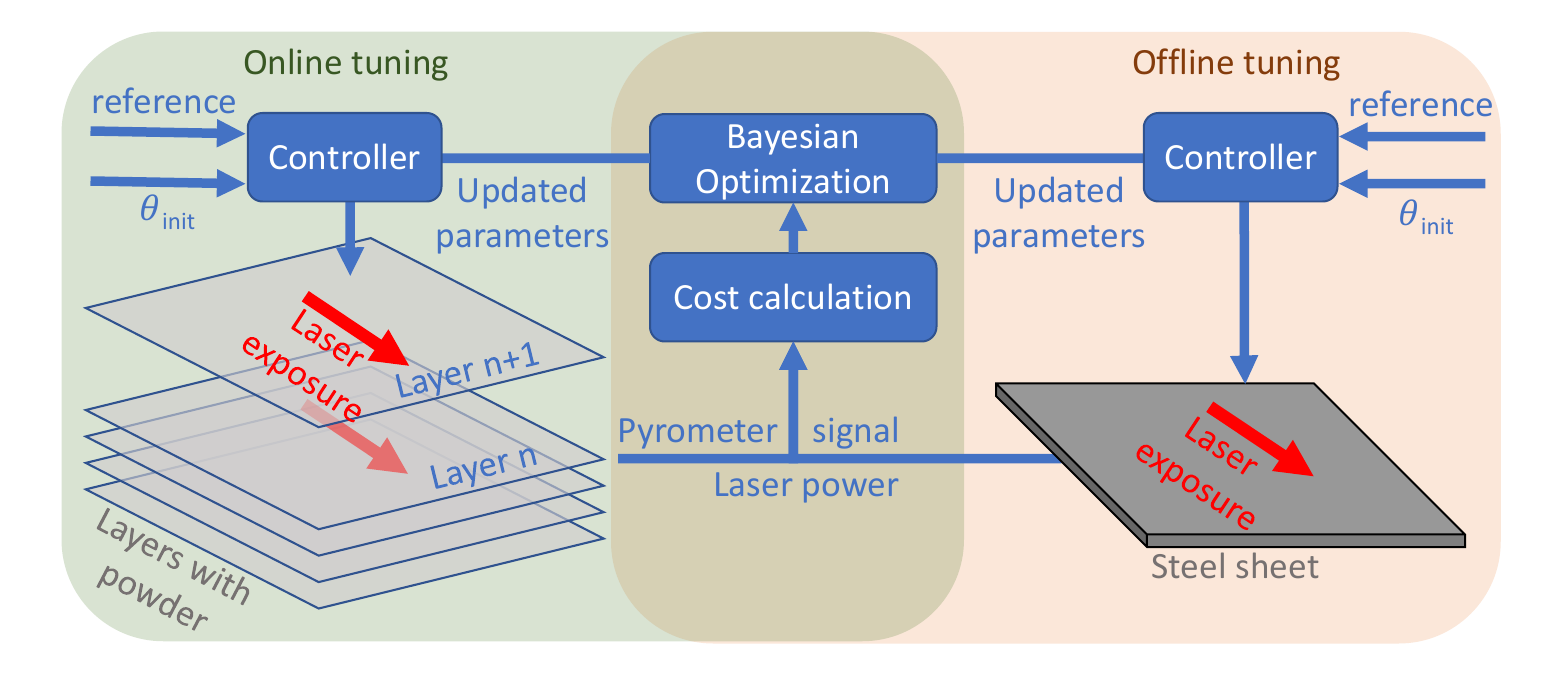}
\caption{Schematic for online and offline controller tuning procedures}
\label{fig:online_vs_offline}
\end{center}
\end{figure}

The proposed auto-tuning procedure is evaluated here for \textit{online} and \textit{offline} conditions.
Schematic representation of the procedures are shown in \Cref{fig:online_vs_offline}.

Offline tuning involves exposing a bulky metal object, i.e., a steel plate instead of a powder layer, to evaluate controller performance in each iteration.
The laser is directed at the same region during each iteration after updating the controller parameters.
The controller parameters that minimize the cost function by the end of the designated number of iterations are applied during the build job to the parts of interest.
This procedure would be performed prior to starting a build job with powder by marking the build plate.
It is a simplified setting compared to an actual print job with powder since the melt pool emissions from a solid substrate exposure are expected to be more stable.
Furthermore, no powder is consumed by this tuning method and iterations are more rapid since no time is required to recoat.
However, this method may be less representative of the actual process conditions.

Online controller tuning is performed at the beginning of the printing process with powder and can be done in locations outside the actual parts to be built.
In this case, a single-vector thin-walled geometry is added to the build job. The wall is exposed and recoated with powder to complete a single iteration.
Meanwhile, the parts are printed without control using standard fixed parameters until the tuning is finished.
After finalizing the tuning procedure, the controller parameters with the minimum cost are applied for controlling the parts in the remaining layers.
The melt pool emissions from the single-vector wall tend to be less stable in the presence of powder.
The build-up of material over several layers also represents a commitment to building a part, or at a minimum, to removing the wall from the build plate before it can be reused.
However, this method is more representative of actual process conditions than offline tuning.

\subsection{Experiment design}

\begin{figure}[ht!]
\begin{center}
\includegraphics[width=0.45\columnwidth]{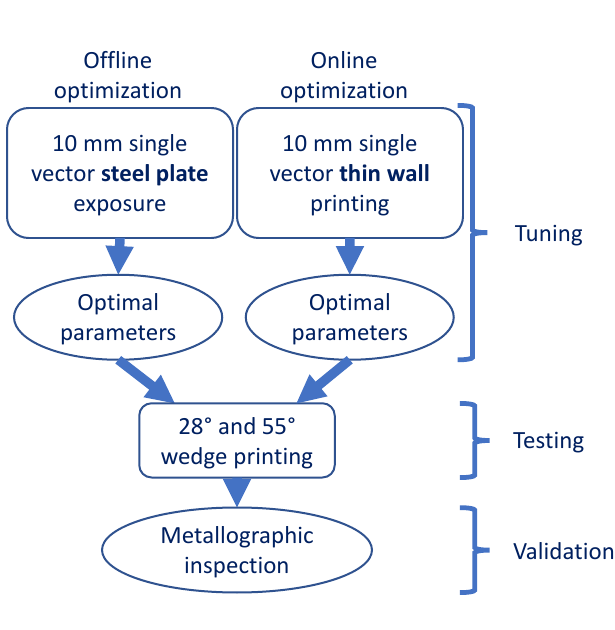}
\caption{Schematic of the experimental plan}
\label{fig:experiment_plan}
\end{center}
\end{figure}

Two experiments were designed for the implementations of the offline and online optimization methods as shown in \Cref{fig:experiment_plan}.

\begin{figure}[ht!]
\begin{center}
\includegraphics[width=0.8\columnwidth]{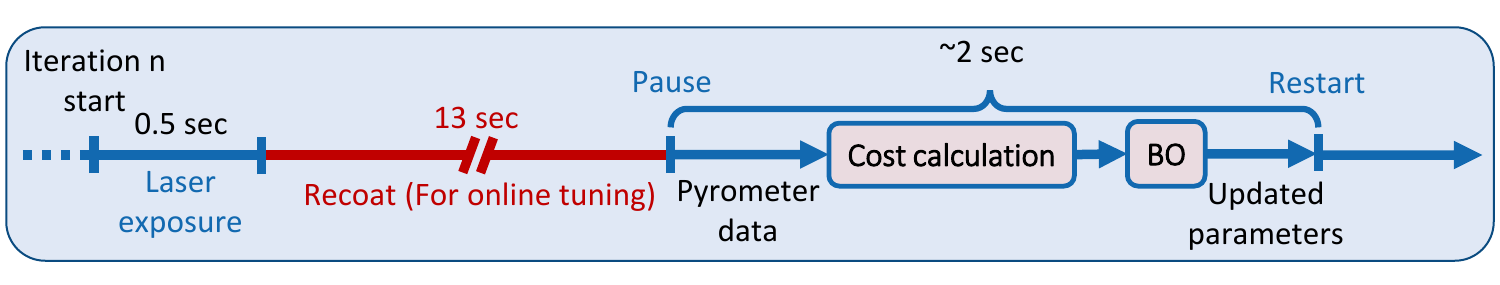}
\caption{Single iteration implementation timeline}
\label{fig:online_offline_schedule}
\end{center}
\end{figure}

A 3mm-thick S304 steel plate was exposed with the 10mm-long single vector for offline optimization for 200 iterations.
A build job with S316L powder was performed for 200 layers by performing the online optimization cycle on a vector in powder with layer applications.
The optimal controller parameter sets were acquired from online and offline tuning procedures.
The timing sequence of the iteration execution is described both for the offline and online settings in \Cref{fig:online_offline_schedule}.

An iteration begins with the exposure of a single line scan of 10mm with the PI controller running and with the input power and output pyrometer signal being recorded.
Due to the melt pool emission difference between plate and powder, applying the same reference value for the PI controller yields a different range of laser power inputs.
To obtain the laser power application in the same range, a reference value of 60 mV is assigned for the online controller tuning instead of 30 mV for the offline.
For the online tuning, recoating is performed after the laser exposure, which takes 13 seconds using the machine's standard settings.
The build process is then paused while the pyrometer signal is processed through the cost function, which in turn is fed into the BO algorithm.
The BO algorithm calculates the controller parameters for the next iteration and the parameters are updated before resuming the build process.
The pause to the restart duration takes $\sim2$ seconds, however, most of this time is used by the in-machine network and for the update of the parameters.
The cost calculation and the BO execution themselves take less than 0.2 seconds in total.
For the build job scheduling along with the iteration to iteration parameter update, the Python-based build job execution capabilities of the Acontiy Midi+ machine were used as described in previous work~\cite{kavas2023layer}. 

\begin{figure}[ht!]
\begin{center}
\includegraphics[width=1\columnwidth]{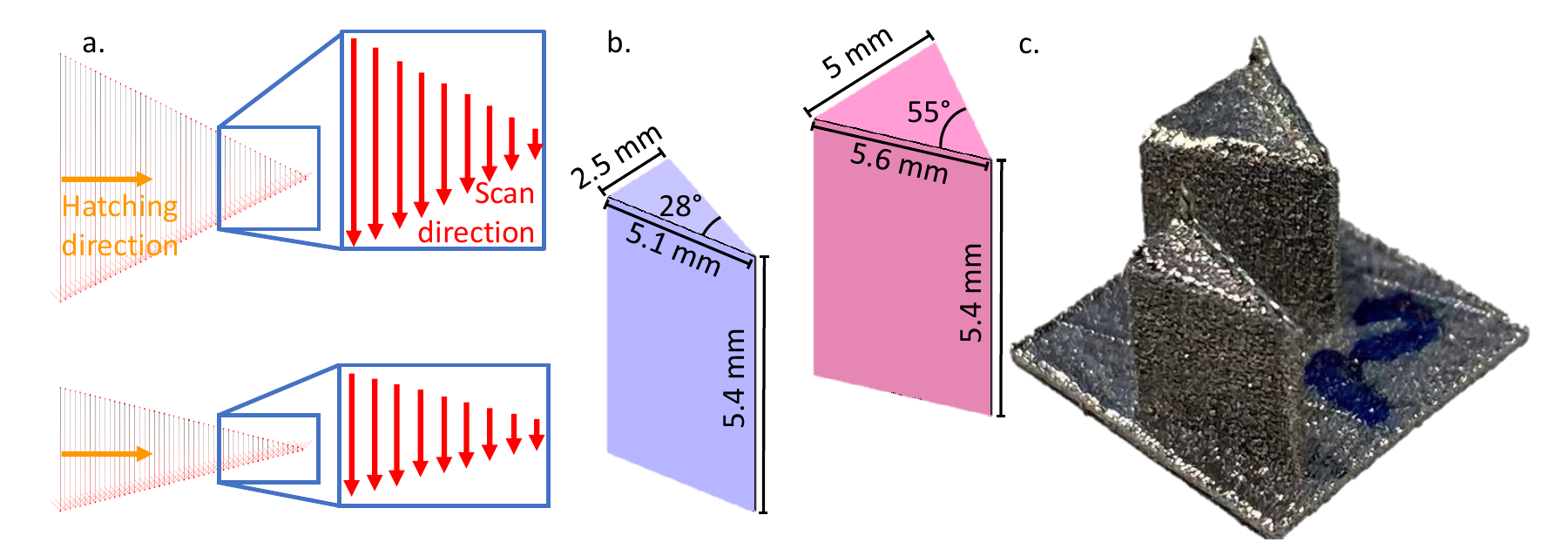}
\caption{Wedge specimen geometries: a) hatched laser vector directions, b) isometric view of the wedges, c) printed geometries}
\label{fig:wedge_geometries}
\end{center}
\end{figure}

For the second part of the experiment that is designed to evaluate the tuned controller performances, two wedge geometries with 28° and 50° angles (as shown in \Cref{fig:wedge_geometries}) were printed.

Due to the pyrometer signal's dependency on the orientation of the laser exposure vector, the scan files were generated using unidirectional scanning, as shown by the vectors with arrows in \Cref{fig:wedge_geometries} (a).
The angles of the wedges were selected arbitrarily to exhibit an actual in-layer heat build-up scenario where the vector-to-vector time gradually becomes smaller.
In total, two sets of the wedge geometries were printed with online and offline tuned controller parameters.

\subsubsection{Microscopic Evaluation}
Specimens were separated from the build platform by electrode-discharge machining. 
They were cut close to their center line by a Struers Accutom-10 cutting machine, then metallographically prepared by embedding, grinding, and polishing.
Embedding is performed hot by a Struers CitoPress machine with DuroLite bakelite resin, then ground with 320-grit sandpaper, and polished with Struers commercial polishing cloths Largo, Dac, Nap, and Chem with suspensions with particle sizes of $\SI{9}{\micro\meter}$, $\SI{3}{\micro\meter}$, $\SI{1}{\micro\meter}$, and $\SI{0.1}{\micro\meter}$, respectively.
Polished mounts were scanned by a Keyence VHX-7000 microscope in coaxial and ring lighting modes.

\section{Theory}
\label{sec:theory}
In this section, the PI laser power controller, pyrometer as the sensor, cost function, and Bayesian Optimization (BO) algorithm are explained in detail.

\subsection{PI Controller}
\label{sec:PID_controller}
The expression 

\begin{align}
    \label{eq:PI_ctrl}
    \text G_{c1}(s) = K_p + \frac{K_I}{s}
\end{align}

represents the transfer function of a PI controller in the Laplace transform domain. 
Here, \( s \) is the complex variable in the Laplace domain, which simplifies the analysis of linear time-invariant systems.
The coefficients \( K_p \) and \( K_I \) stand for the proportional and integral gains, respectively. 
The proportional term \( K_p \) provides a control action proportional to the error. 
On the other hand, the integral term \( \frac{K_I}{s} \) integrates the error over time, aiming to eliminate steady-state error which allows tracking of the reference value ~\cite{bishop2011modern}.

\subsection{Cost function}
\label{sec:cost_function}

A cost function is formulated to quantify the controller performance through the pyrometer signal.
The multiple objectives are identified for quantifying the deviation in the melt pool conditions from the nominal response.
The pyrometer signal is expected to:

\begin{itemize}
    \item track the given reference value with a minimal amount of deviation through a single vector,
    \item reach the reference value in the shortest time possible,
    \item stabilize without oscillation in the steady state.
\end{itemize}

The first criterion conforms to the ideal size of the melt pool to be maintained through the hatch region, regardless of the varying preheating conditions, vector lengths, or defects.
The second criterion serves to shorten the time of melt pool transition time to stabilize at the initiation and is referred to as the rise time. 
The third criterion is introduced for limiting the controller-induced oscillations, especially for larger controller gains.

For these criteria, a composite cost function term $g$ is formulated for the optimization problem is given as

\begin{equation}
    \label{eq:error}
    g(x) = \sqrt{MSE'^2 + t_{\mathrm{rise}}'^2 + \sigma_{laser}'^2}
\end{equation}



where $MSE'$, $t_{\mathrm{rise}}'$, and $\sigma_{laser}'$ denote the normalized mean squared error,  normalized rise time, and normalized standard deviation, respectively.
These can be combined into a single error metric, $g$, using the Euclidean norm where $[MSE', t_{\mathrm{rise}}', \sigma_{laser}']^T$ is the error vector.

\begin{figure}[ht!]
\begin{center}
\includegraphics[width=1\columnwidth]{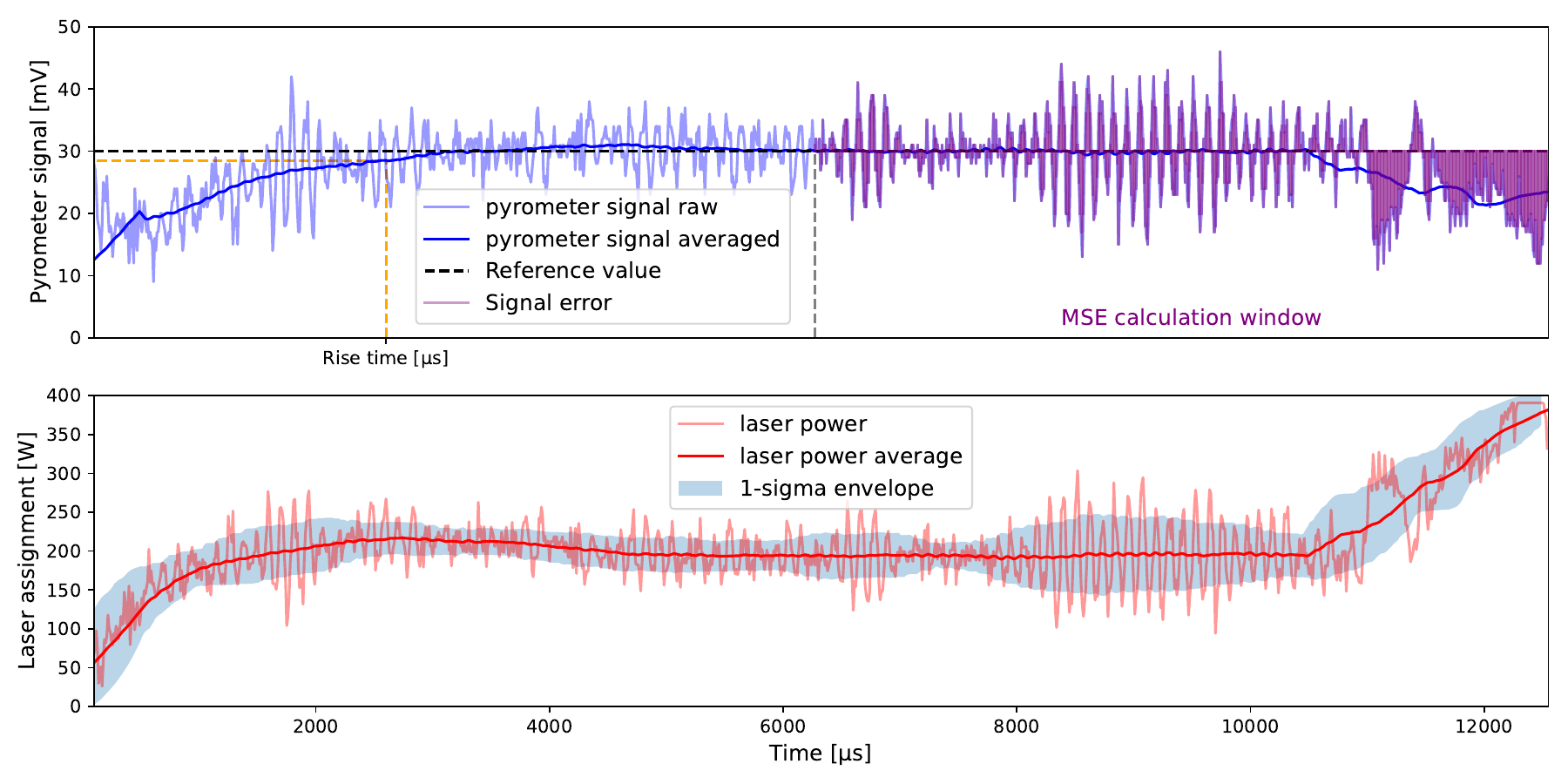}
\caption{Description of the three cost function terms in the pyrometer and the laser power input signals in time series for an exemplary laser strike. 
Signal averages are visualized with zero-phase shift using a moving window of size 100.}
\label{fig:error_function}
\end{center}
\end{figure}

An illustrative description of the components of the proposed cost function applied to a typical laser strike is shown in \Cref{fig:error_function}.
Next, each term of the cost function is explained in detail.

First, normalized mean squared error ($MSE'$) is formulated as

\begin{align}
    \label{eq:mse}
    \text{$MSE'(y_i)$} = \frac{1}{Nc_{MSE}} \sum_{i=N/2}^{N} (y_i - \hat{y}_i)^2,
\end{align}

where $y_i$ is the pyrometer measurement and $\hat{y}_i$ is the reference value.
$N$ is the number of the recorded pyrometer data, present in the denominator for normalizing the length of the exposed vector. 
This term is used to penalize the deviation of the pyrometer signal from the set point.
The term is only calculated for the second half of the vector (last 5mm) to be less affected by the rise time as shown by the purple-colored signal error lines in the second half of the first plot in \Cref{fig:error_function}.
Given the significant disparity in magnitude between units, such as laser power in the order of hundreds of Watts and pyrometer signals in the tens-of-mV range, each term's impact on the composite error varies considerably.
To balance the weight of each term, a second normalization factor $c_{MSE}$ is introduced to be multiplied with the vector length $N$.
For the $MSE$ term calculation, $c_{MSE}$ is chosen as 500.

Second, the rise time is defined as
\begin{align}
    \label{eq:rise_time}
    t_{\mathrm{rise}}'(y_i) = 
    \begin{cases} 
    1 \quad \text{if } \frac{1}{N}\arg\min_k \left\{k: \big| y_i -\hat{y}_i\big| \leq 0.05\hat{y}_i\right \} = 0, \\
    \frac{1}{N}\arg\min_k \left\{k: \big| y_i -\hat{y}_i\big| \leq 0.05\hat{y}_i\right \} \quad \text{otherwise}.
    \end{cases}
\end{align}


where $N$, $y$, and $\hat{y}_i$ are the same as Equation \ref{eq:mse} and $k$ represents the number of data points before reaching the 95\% of the reference value as shown by the orange dashed line in \Cref{fig:error_function}.
If the signal fails to reach the target completely, i.e. $k=0$, the rise time term yields 1.
Shorter rise times are thus favored by this cost function term.

Lastly, the standard deviation is formulated as

\begin{align}
    \label{eq:std}
    \sigma_{laser}'(P) = \frac{\sqrt{\frac{1}{N} \sum_{i=1}^{N} (p_i - \mu_{i})^2}}{c_{\sigma}}  \, ,
\end{align}

where $p$ is the laser power assignment and $\mu$ defines the rolling average of the laser power assignments. 
For the $\sigma_{laser}$ term calculation, $c_{\sigma}$ is chosen to be 150.
$\mu$ is given by

\begin{align}
    \label{eq:mean}
    \mu_{i} = \frac{1}{w} \sum_{j=i}^{i+w-1} y_{j} \, ,
\end{align}

where $w$ is the window length, defined to be 100 for the experiments of this study.
Unlike the first two terms, the standard deviation is calculated for the laser power input instead of the pyrometer signal.
Due to the steady-state oscillation of the pyrometer signal that can be characterized by the combined effect of the process physics and sensor noise, the normalized standard deviation of the laser power signal is used to isolate the controller-induced oscillations.
This term of the cost function favors the stability of the process input.

\begin{figure}[ht!]
\begin{center}
\includegraphics[width=1\columnwidth]{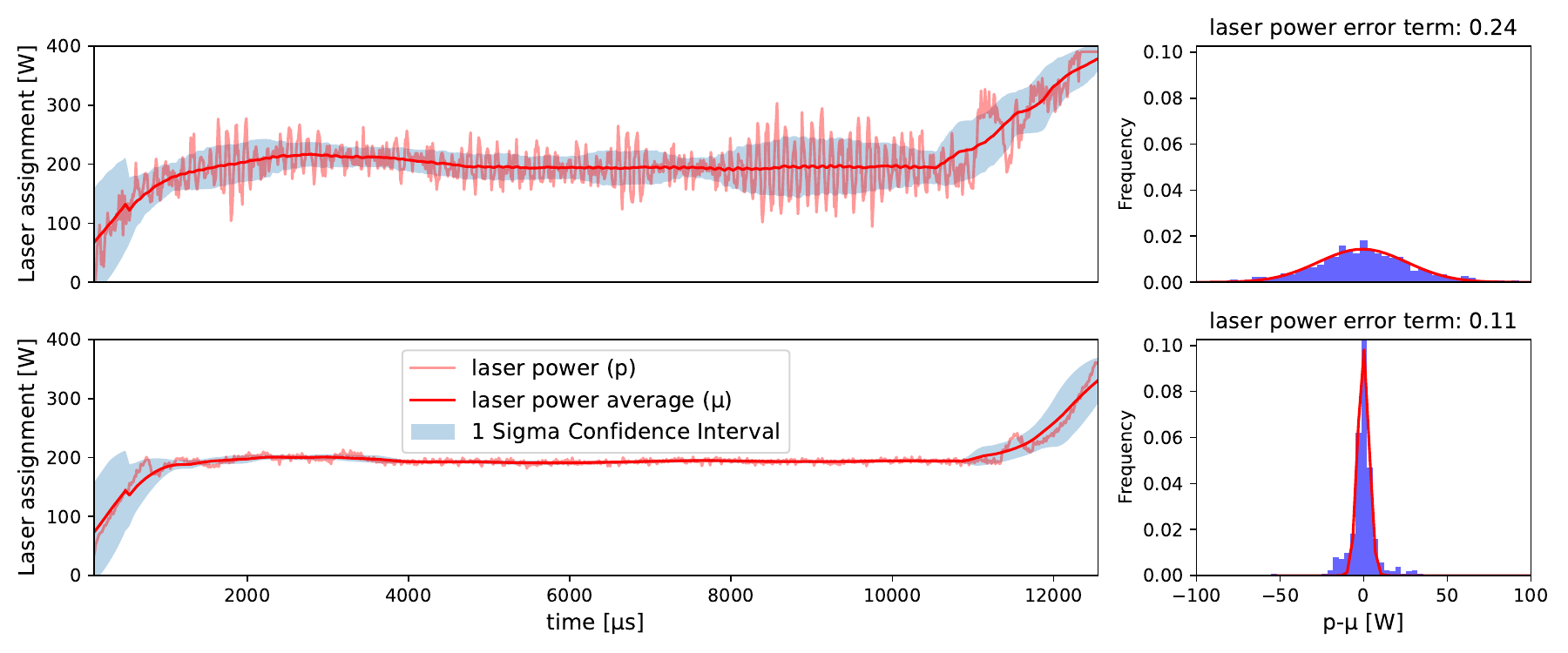}
\caption{Examples of poorly tuned (top) and well-tuned PI controller parameters for an exemplary laser strike.
}

\label{fig:laser_power_good_bad_comparison}
\end{center}
\end{figure}

Examples of low and high-cost laser power signals are given in \Cref{fig:laser_power_good_bad_comparison}.
The first iteration example has high magnitude oscillation in the laser power that is induced by the controller.
The histogram visualization of the laser power assignments subtracted from the rolling average mean exhibits a broader distribution of the laser power assignments that yields a normalized standard deviation value of 0.24 for the entire vector.
The second iteration example has a lower normalized standard deviation value of 0.11 as it does not exhibit an oscillatory behavior.

\subsection{Bayesian Optimization (BO)}
\label{sec:BO}

To obtain the optimal $x$ based on the defined cost function $g(x)$ while preserving stability in the system, we need to solve the following optimization problem:

\begin{equation}
    \label{eq:optimization}
    \min_{x \in A} g(x),
\end{equation}

subject to the constraints

\begin{equation}
    \label{eq:optimization_limits}
    f_j(x) \leq J_{\text{max}}, \quad \forall j \in \{1, 2, \ldots, J\},
\end{equation}

where \( x \) is a decision variable vector within a continuous search space \( A \subset \mathbb{R}^n \). 
Here, \( J_{\text{max}} \) is the pre-set constraint limit indicating the assignment limits of $K_p$ and $K_i$, \( g: \mathbb{R}^n \to \mathbb{R} \) is the objective function to minimize, and \( f_j: \mathbb{R}^n \to \mathbb{R}, j = 1, \ldots, J \) are constraints that must be fulfilled. 
In this study, the range of the controller parameters is only limited by the design limits of the experimental setup, i.e., supported by the AconityCONTROL software.
The functional forms of \( g \), for \( j = 0, \ldots, J \), are unknown, but measurements of \( g \) can be obtained to develop surrogate models using Gaussian Processes (GPs).
The inputs ($K_p$ and $K_i$) and the output as the cost function ($g$) are defined as GPs for the minimization objective defined in \ref{eq:optimization}.

Following the methodology of Berkenkamp et al.~\cite{berkenkamp2016safe}, GPs are used for approximating \( g_j \), for \( j = 0, \ldots, J \).
The approximations are \( \tilde{g}_j(x): A \to \mathbb{R} \), where \( j = 0 \) is for the objective function (1a), and \( j = 1, \ldots, J \) are for the constraints. 
Gaussian process regression presumes that the values \( \tilde{g}(x_0), \tilde{g}(x_1), \ldots, \tilde{g}(x_P) \) for different \( x \) are random variables with a joint Gaussian distribution for any finite \( P \). The model of \( \tilde{g}_j \) is a GP, which is characterized by known mean \( \psi_j(\cdot) \) and kernel \( k_j(\cdot, \cdot) \) functions:

\begin{equation}
    \label{eq:gp}
    \tilde{g}_j(x) \sim \text{GP}(\psi_j(x), k_j(x, x)),
\end{equation}

where the kernel function used in this study is the Radial Basis Function (RBF), described as

\begin{equation}
    k(x, x') = \sigma^2 \exp\left(-\frac{\|x - x'\|^2}{2l^2}\right) \, ,
\end{equation}

where $\sigma^2$ is the variance and $l$ is the length scale
$\sigma^2$ and $l$ are the hyperparameters, described in \Cref{sec:hyperparameter_selection} below.

In this setting, the usual assumption is having access to noisy measurements \( \hat{g}_j(x) = g_j(x) + \omega \), where \( \omega \sim \mathcal{N}(0, \sigma^2_{\omega}) \) is a zero mean normally distributed random variable with $\sigma_{\omega}$ standard deviation. 
To integrate these GPs into optimization, value of \( \tilde{g}_j \) is predicted at a random point \( \hat{x} \) using \( R \) past measurement data \( G_j = [\hat{g}_j(x_r)]_{r=1,\ldots,R} \). 
As per Rasmussen and Williams~\cite{rasmussen2006gaussian}, the mean and variance of the prediction at a new point \( \hat{x} \) are:

\begin{equation}
    \mu_j(\hat{x}) = \psi_j(\hat{x}) + k_R(\hat{x})(K_R + I_R\sigma^2_{\omega})^{-1}(G_j - \Psi_j),
\end{equation}

\begin{equation}
    \sigma^2_{R,j}(\hat{x}) = k(\hat{x}, \hat{x}) - k_R(\hat{x})(K_R + I_R\sigma^2_{\omega})^{-1}k^T_R(\hat{x}),
\end{equation}

where \( G_j \) is a vector of \( R \) observed noisy values, \( \Psi_j = [\psi_j(x_r)]_{r=1,\ldots,R} \) is a vector of mean values of the past data, \( j = 0, \ldots, J \), \( K_R \) is the covariance matrix of past data, \( k(x_a, x_b), a, b = 1, \ldots, R \), \( k_R(\hat{x}) \) contains the covariance between the new point and the past data, and \( I_R \) is the identity matrix of dimension \( R \).

The acquisition function \( acq: X \rightarrow \mathbb{R} \) is employed to evaluate the next sample point in the next iteration based on the criteria of \textit{highest standard deviation}. 
In this work, we use an acquisition function corresponding to the Lower Confidence Bound (LCB), with a modification to ensure more efficient computation under safety constraints, following \cite{zagorowska2023efficient}. 
Thus, at each iteration we evaluate and optimize the acquisition function on the corresponding GP posterior as follows 

\begin{equation}
    x_{j+1} = \arg\max_x acq(x).  
\end{equation}


\subsubsection{Hyperparameter selection}
\label{sec:hyperparameter_selection}
There are 3 hyperparameters tuned for the used algorithm: initialization for P and I, variance $\sigma^2$, and length $l$ scale of the kernel function.
The variance $\sigma^2$ sets the overall amplitude of the function, with a higher variance allowing for greater fluctuations from the mean.
The length scale $l$ determines the smoothness of the predicted function, where a smaller $l$ leads to rapid variations in the function, and a larger $l$ results in smoother behavior. 
In dynamic system modeling using GPs, hyperparameters are typically estimated via maximum likelihood based on observed data~\cite{ostafew2016robust}.
However, in BO, the GP model not only performs regression on existing data but also actively requests new samples through data acquisition.
Therefore, hyperparameters are set before initialization and maintained through the BO procedure, treating the kernel as a prior over functions.

In this study, $\sigma^2$ is set to 1.
$l$ of the GP is set as 50 and 100 for the inputs P, and I, and 200 for the cost as output based on the minimization of the negative log-likelihood for a training dataset. 
Initial values are set to the lower bound values for both P and I as 1 and 100, respectively.
These values are assigned based on the historical experiments.
A detailed study on the effects of hyperparameters on tuning and the setting of the initial parameter values is beyond the scope of this study.
The limits are set to 1 to 100 for $K_p$ and 100 to 1,600,000 for $K_i$ for the BO algorithm as the upper and lower boundaries are dictated by the machine interface.
During the iterations, the range and order of magnitude of the controller parameters susceptible to disruptive instability are unknown.
Input parameter restriction by utilizing safe-BO algorithms can be employed to avoid these cases during the optimization~\cite{khosravi2022safety}.
For this study, no such limit is issued besides the present hardware limits.
The safe progression of the iterations is ensured by assigning an upper limit for the laser power assignment of 300W, which was found during separate experiments to be within the stable process window for the studied material with the process parameters described above.
Although instability may be observed during iterations, excessive melting is avoided by the laser power limit by ensuring a safe process for any suboptimal controller setting.


\section{Results}
\label{sec:results}

\subsection{Optimization}

\begin{figure}[ht!]
\begin{center}
\includegraphics[width=0.7\columnwidth]{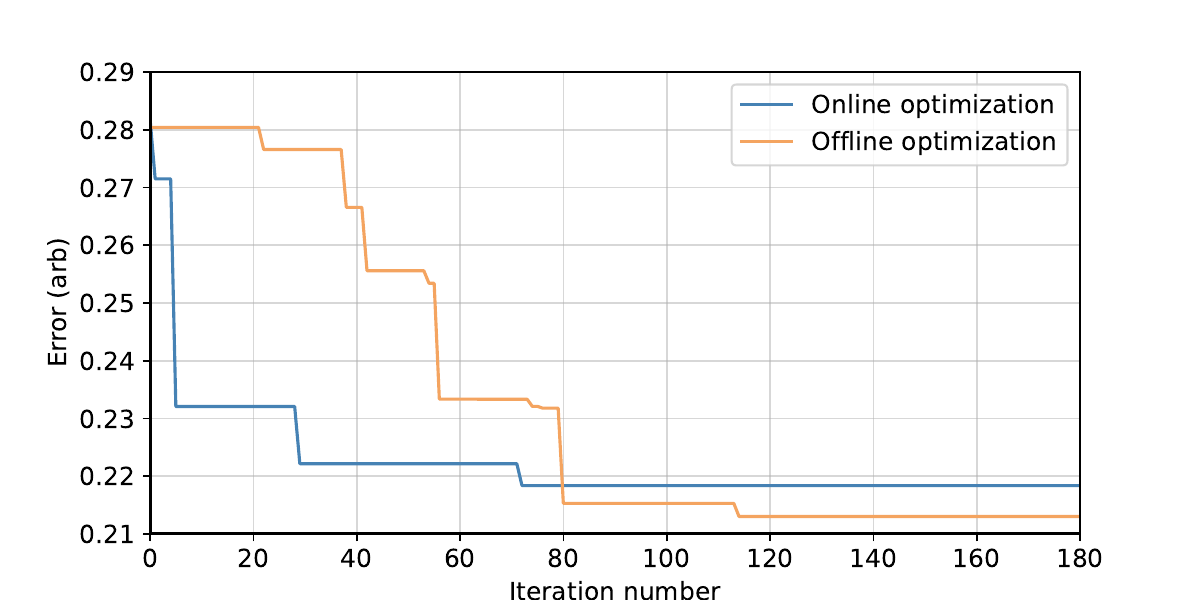}
\caption{Minimum achieved cost function error as a function of iteration number for the online and offline tuning experiments.}
\label{fig:optimization_iterations}
\end{center}
\end{figure}

Optimization progress is shown in \Cref{fig:optimization_iterations} for both online and offline settings.
The line plots represent the lowest achieved error value of each previous iteration.

The minimum cost function value is reached at iteration number 112 for the offline setting and iteration number 74 for the online setting.
Online optimization reached the minimum earlier than the offline settings with a faster drop in the error value.
Offline optimization 
The offline optimized values are $P=8.45$ and $I=90598.24$. 
Meanwhile, the online optimization yielded $P=8.44$ and $I=65014.97$.

\begin{figure}[ht!]
\begin{center}
\includegraphics[width=1\columnwidth]{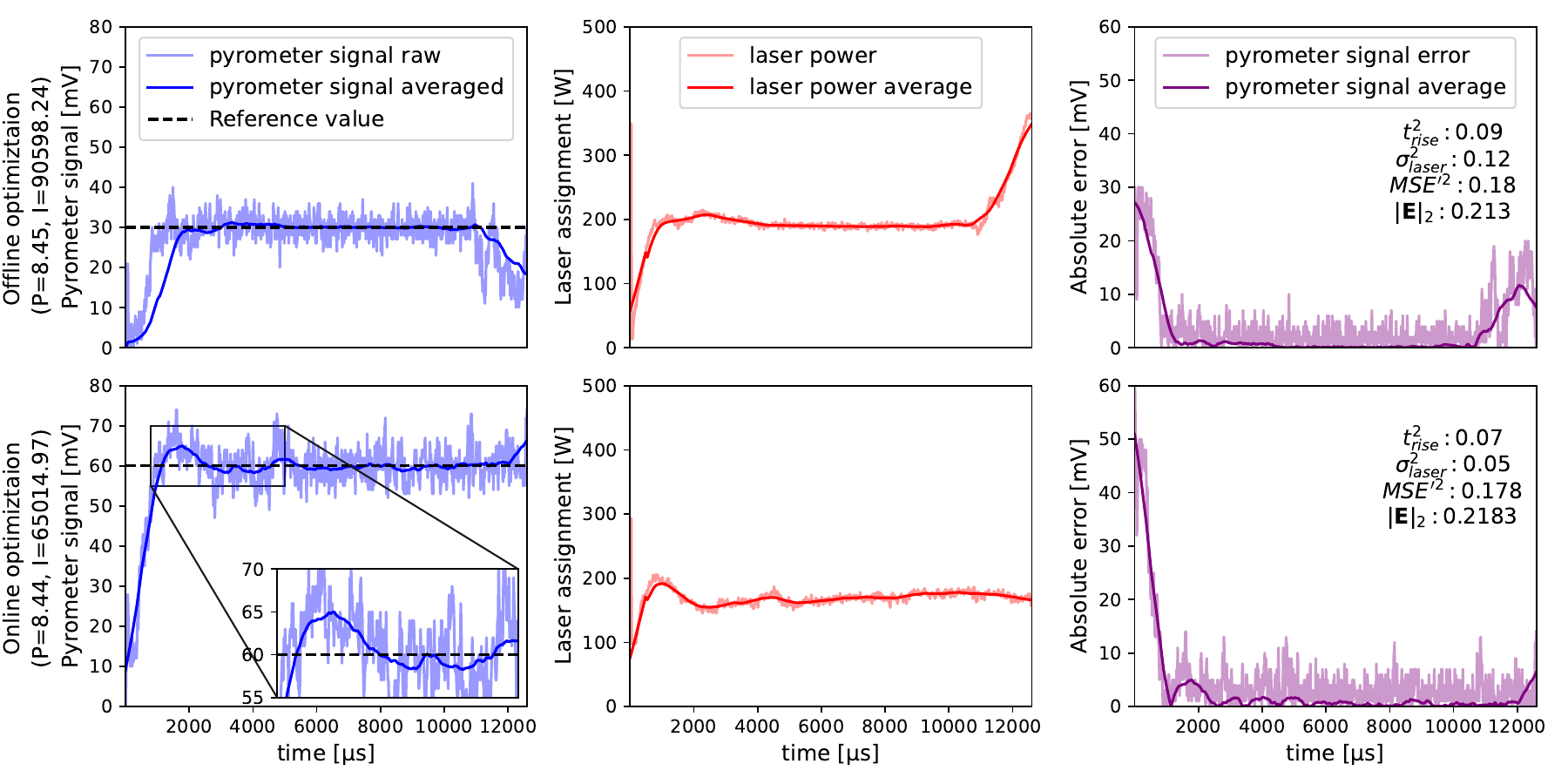}
\caption{The pyrometer signal, assigned laser power, and the error of the lowest cost function value iterations of the offline (top row) and online (bottom row) optimization iterations. Each plot presents the data from a single vector.}
\label{fig:best_iterations_combined}
\end{center}
\end{figure}

Respective time series plots of the pyrometer signal, laser power assignment, and error plots are shown for a single vector with the optimized PI settings in \Cref{fig:best_iterations_combined}. 

The first row displays the optimal iteration from the offline optimization process, while the second row presents the same from the online optimization. 
The difference in the reference pyrometer values is achieved in a similar laser power range of around 200W due to the emission difference of the melt pools in steel plate and on powder.
Overshooting is observed in the online optimization case, unlike the offline one shown closely in the magnified inset plot.

\subsection{Wedge prints}

\begin{figure}[ht!]
\begin{center}
\includegraphics[width=1\columnwidth]{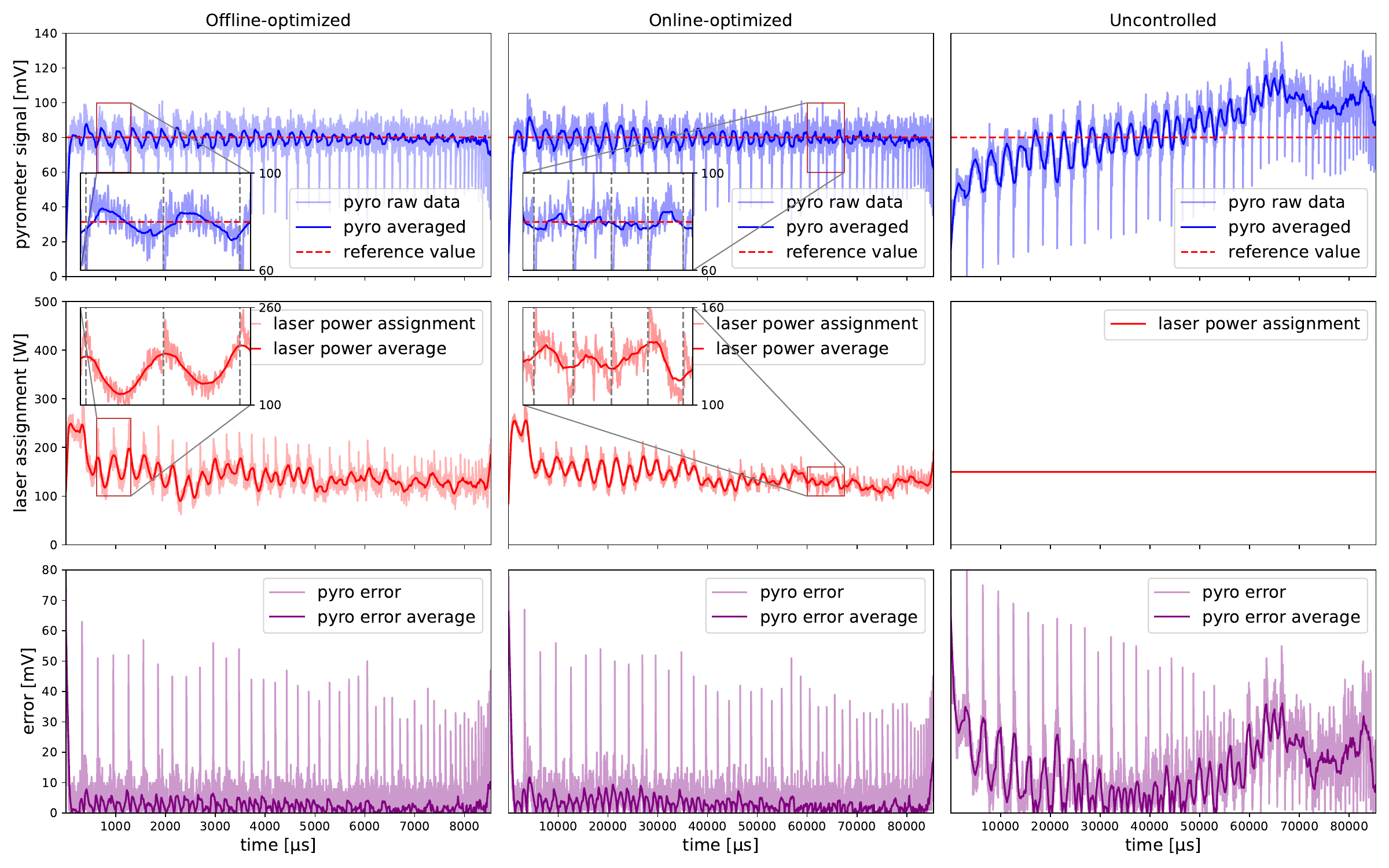}
\caption{Online-optimized PI-controlled (left column), offline-optimized PI-controlled (center column), and uncontrolled (right column) time series signal comparison of the 28° wedge specimens at layer 90. The rolling average is performed with a window size of 100 data points.}
\label{fig:wedge_28_timeseries}
\end{center}
\end{figure}

\begin{figure}[ht!]
\begin{center}
\includegraphics[width=1\columnwidth]{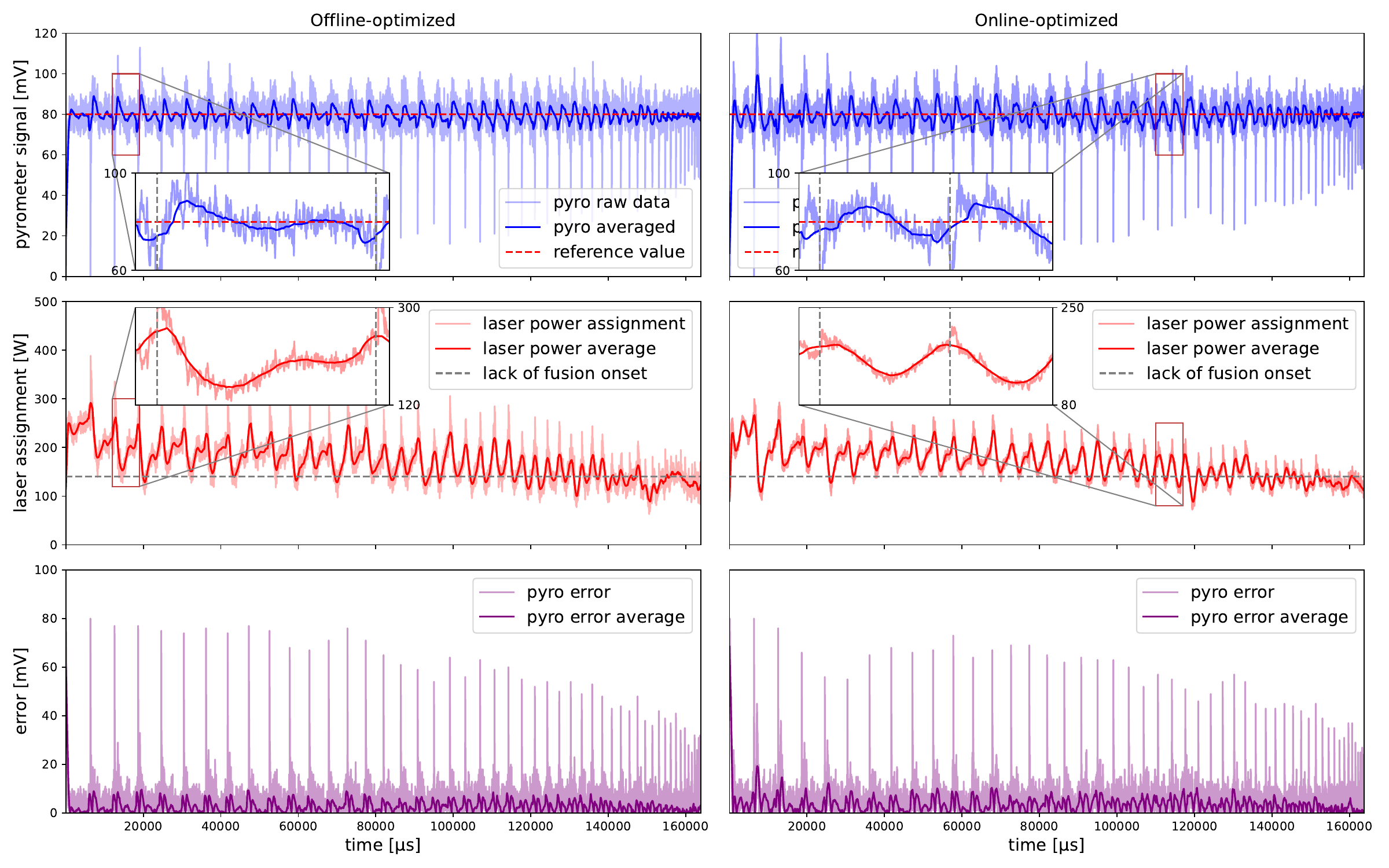}
\caption{Online-optimized PI-controlled (left column), offline-optimized PI-controlled (right column) of 55° wedge specimens at layer 90. The rolling average is performed with a window size of 100 data points.}
\label{fig:wedge_55_timeseries}
\end{center}
\end{figure}

Time series and scatter plots of the 28° and 55° wedge parts are shown in \Cref{fig:wedge_28_timeseries} and \Cref{fig:wedge_55_timeseries}, respectively.
The inter-vector travel time of the scanner has been removed from the time-series visualization of the wedge prints.

The uncontrolled wedge print is only performed for the 28° wedge part.
In the plot of the 28° wedge part, the magnified inset plots are shown for the pyrometer signals and the laser power input for the controlled wedge parts and the vector start time points are shown by the vertical dashed grey lines.
After an initial transient, the pyrometer signal for both the online- and offline-tuned PI-controlled 28° wedges is seen to stabilize about the set point value for the majority of the hatch vectors along the wedge.
However, as it is shown in the inset plots, overshooting followed by overcompensating is omnipresent in the first half of the hatch region.
The overshooting can be attributed to the controller behavior as the causal trend is also observed in the laser power values.
Correspondingly, the laser power assignment is shown to vary along each hatch vector.
Only at very short vector lengths (i.e., the end of the time series data) does the pyrometer signal begin to diverge from the command.
The uncontrolled case shows a distinct heat build-up behavior as a result of getting smaller vectors due to the geometry of the wedge parts.

The 55° wedge part plot shows a vector-to-vector trend similar to the 28° wedge part.
Overshooting is also present for the majority of the vectors, however, the signal settles in the earlier and longer vectors as the inset plots of the pyrometer signal show.
later and shorter vectors resemble the same trend as the initial vectors of the 28° wedge part with a shorter length.
This resemblance can be attributed to the controller behavior's dependency on the vector length.
The overheating compensation is apparent with the laser power values decaying to the lack-of-fusion zone starting at 140W as shown by the grey dashed line.

\begin{figure}[ht!]
\begin{center}
\includegraphics[width=1\columnwidth]{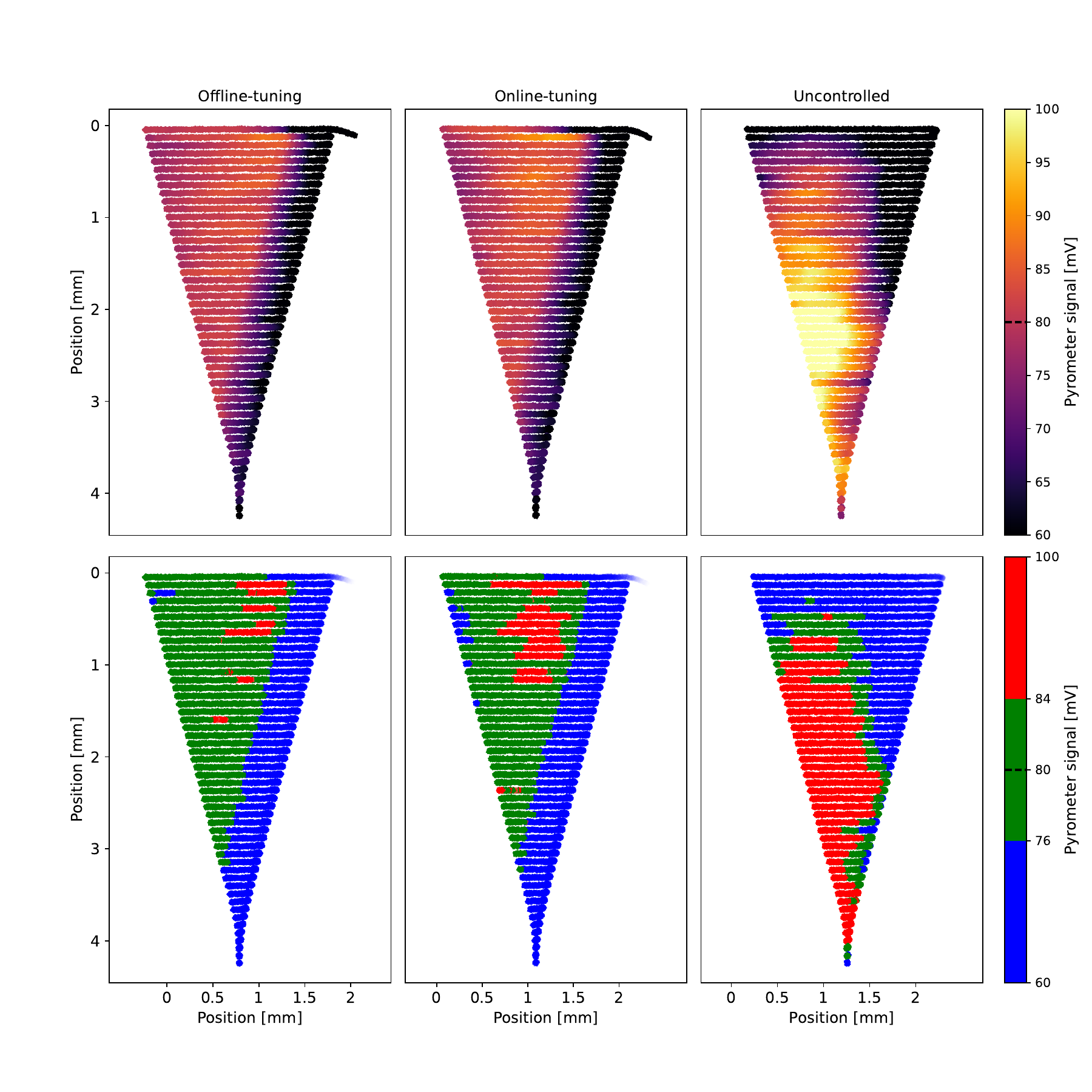}
\caption{Scatter plots of the pyrometer signal of the 28° wedge part at layer 90 for offline- and online-tuned controllers on the left and the right column, respectively. The first row presents the pyrometer signal on a gradient color map. The second row presents the same data with a discretized color map where the green points represent the $\pm 5\%$ range of the reference value 80 which is interpreted as the nominal controller performance. Red and blue points represent higher and lower than the nominal range.}
\label{fig:scatter_28}
\end{center}
\end{figure}

\begin{figure}[ht!]
\begin{center}
\includegraphics[width=1\columnwidth]{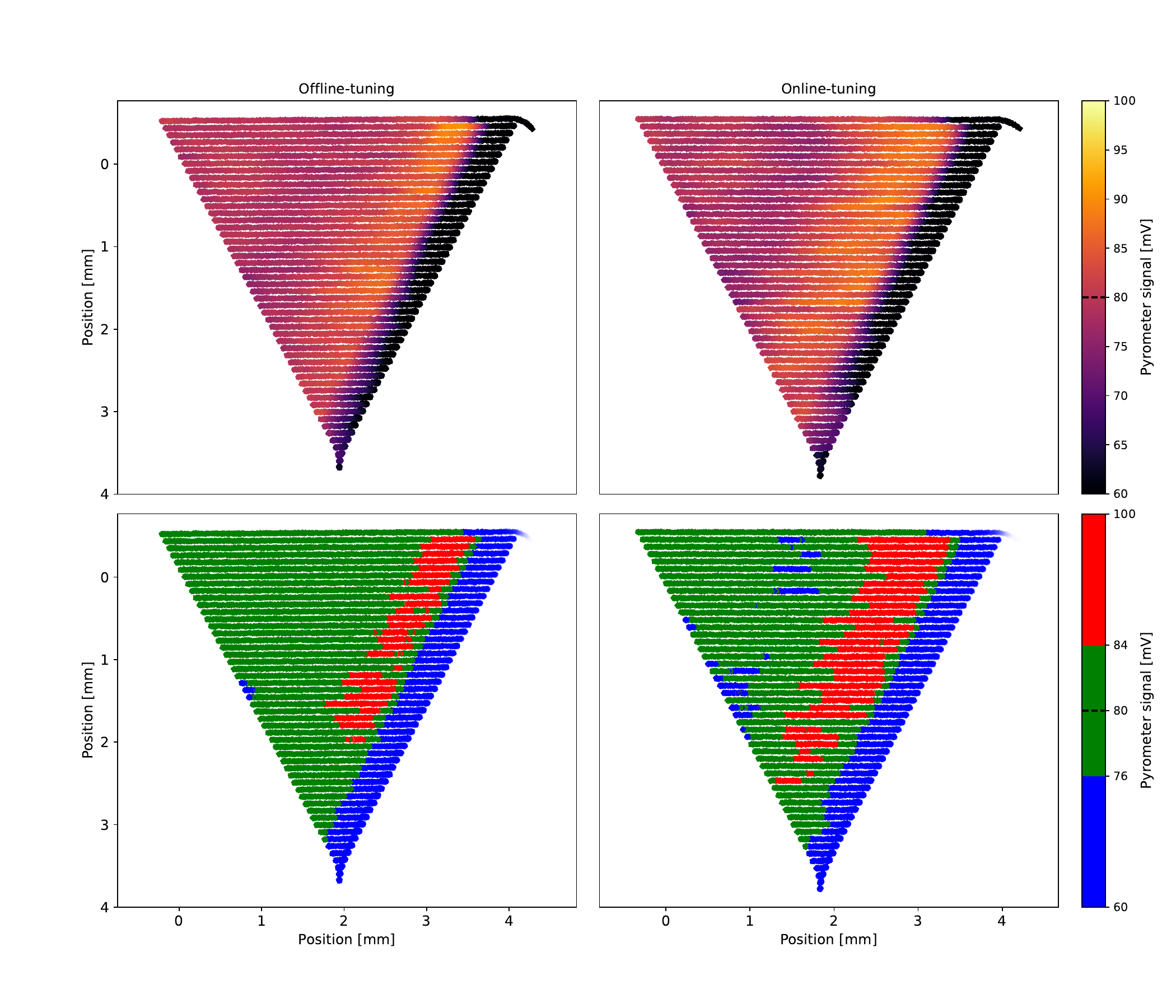}
\caption{Scatter plots of the pyrometer signal of the 55° wedge part at layer 90 for offline- and online-tuned controllers on the left and the right column, respectively. The first row presents the pyrometer signal on a gradient color map. The Second row presents the same data with a discretized color map where the green points represent the $\pm 5\%$ range of the reference value 80 that is interpreted as the nominal controller performance. Red and blue points represent higher and lower than the nominal range.}
\label{fig:scatter_50}
\end{center}
\end{figure}

Spatial plots of the pyrometer signal for the 28° and 55° wedge parts are shown in \Cref{fig:scatter_28} and \Cref{fig:scatter_50}, respectively. 

The top row shows the pyrometer signal of the offline and online tuned controllers resulting pyrometer signal on the 28° wedge part as well as the uncontrolled signal.
The coloring of the presented data is centered around the reference value of 80 mV. 
The bottom rows show the same data with a different coloring to represent the performance of the controller.
The green points represent the $\pm 5\%$ range of the reference value as the reds and blues are above and below this range, respectively.
The pyrometer signal trend observed in the time series is also observable spatially for each vector.
The start of all vectors in the controlled cases experience the rise time duration represented by the blue starting end.
The length of this region is constant through the vectors regardless of the vector length.
It is followed by the overshooting region that is represented by the red regions in the controlled cases, which are only apparent for the longer initial vectors.
Based on the time series plots as shared in \Cref{fig:wedge_28_timeseries} and \Cref{fig:wedge_55_timeseries}, overshooting is observed in all vectors, however, it does not exceed the $+ 5\%$ reference value in shorter vectors.
The uncontrolled part shows a very distinct and asymmetrical overheating that is attributed to the unidirectional exposure of the hatch region.
The green region in the uncontrolled part is merely present as a short transient region where the signal traverses as the part undergoes overheating, compared to the stable vector-to-vector trend of the controlled parts.
In both wedge parts, the online-tuned controller yields less overshooting compared to the offline-tuned controller.

\begin{figure}[ht!]
\begin{center}
\includegraphics[width=1\columnwidth]{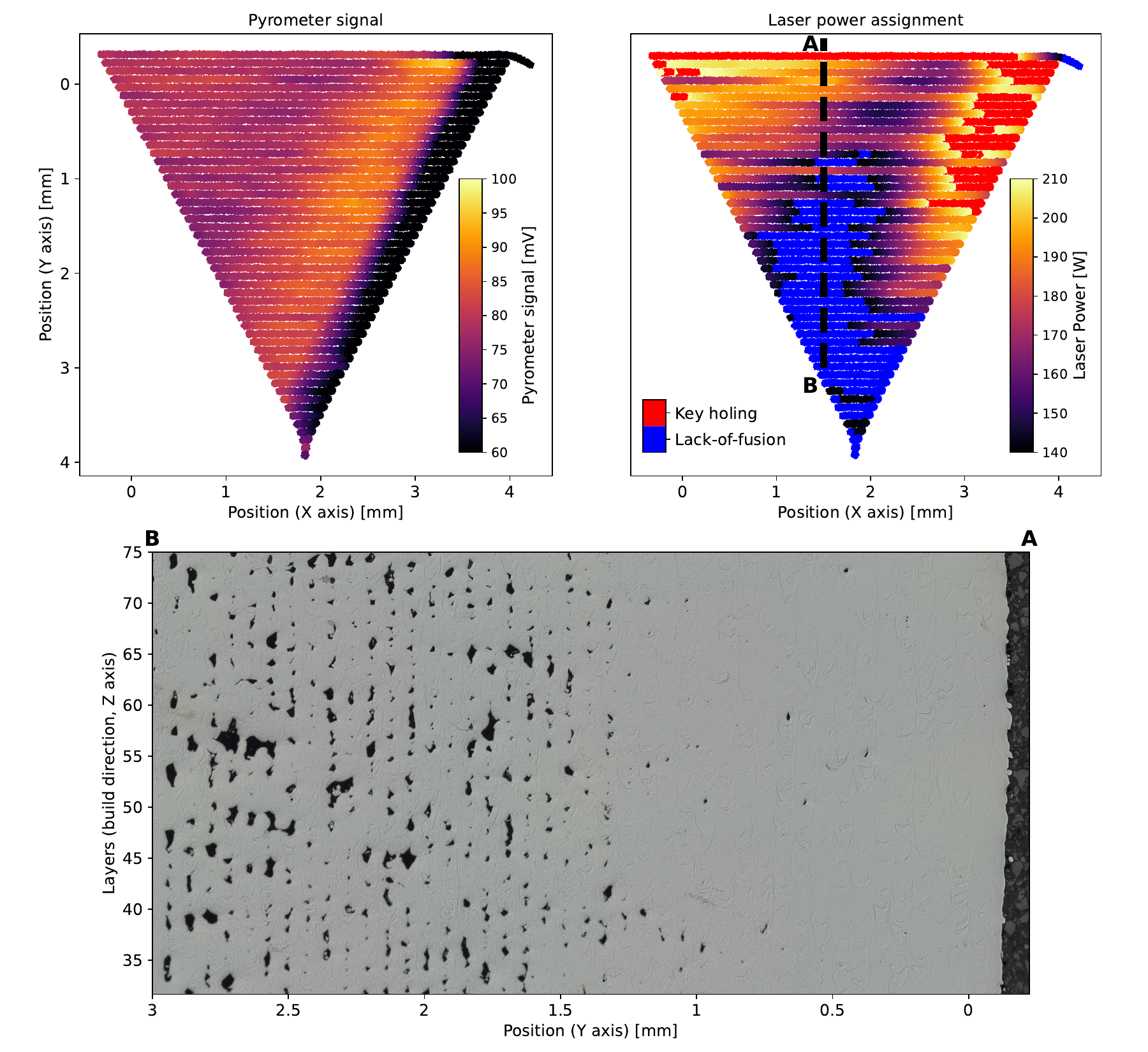}
\caption{Pyrometer signal measurement (upper-left) and the corresponding laser power assignment (upper-right). Laser power assignment is marked with the color map for the nominal process window. Keyhole and lack-of-fusion susceptible parameter ranges are marked with red and blue, respectively. At the bottom, the cut-up image taken through [AB] plane as described in the laser power assignment plot is shown from layers 30 to 75. Hatch progression direction is from A to B (right to left). }
\label{fig:microstructure}
\end{center}
\end{figure}

The pyrometer signal and the corresponding laser power assignment are shared in the first row of \Cref{fig:microstructure}.

The pyrometer signal is visualized by the same color range as in \Cref{fig:scatter_50}.
The laser power assignment is color-mapped between 140W and 210W.
Based on the identified parameter window, it is expected to observe lack-of-fusion and keyhole pores below 140W and above 210W of laser power, respectively.
These porosity expected regions are colored with blue for lack-of-fusion and red for key holing as described by the plot legend.
The metallographic inspection performed through the plane described by AB is shown in the second row.
The microstructural image captures the range of layers from 30 to 75 through the AB plane.
The right side of the image shows the cross-section of the initial vectors in the hatch of every layer and the hatch progresses towards the left side.
The lack-of-fusion type of defects observed on the left side of the image shows a strong correlation with the laser power assignments in the lack-of-fusion range.

\begin{figure}[ht!]
\begin{center}
\includegraphics[width=0.6\columnwidth]{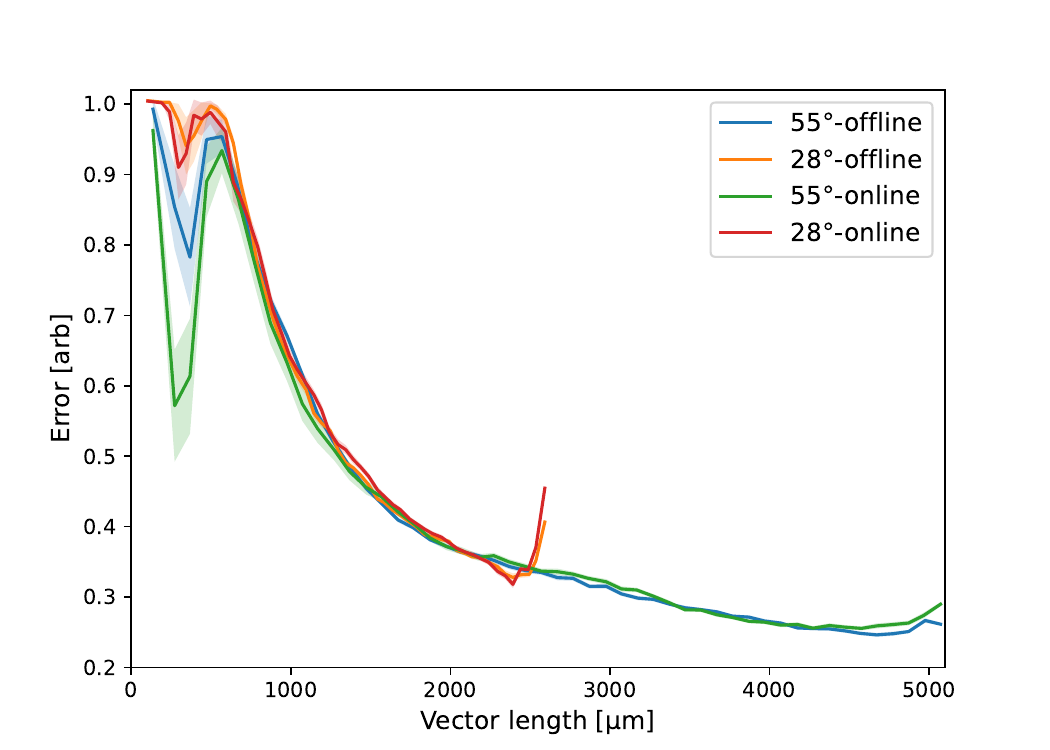}
\caption{Error value per vector length relationship for all printed wedge parts.}
\label{fig:costs_comparison}
\end{center}
\end{figure}

The mean value of the cost function applied to each wedge geometry is shown as a function of vector length for both online- and offline-optimized parameters in \Cref{fig:costs_comparison}.

To compute this, each layer was divided into individual hatch vectors labeled by length.
Then, the cost function used for the optimization was computed for each vector individually.
The same cost calculation equations have been applied to each vector with the same constant values as described in Section \ref{sec:cost_function}.
This procedure was repeated for all layers and the mean of the resulting cost and its terms are plotted against vector length.
The shaded areas represent the 95\% confidence interval of the calculated values about the mean lines.
The overall cost values are observed to be increasing with decreasing vector length.
The trend is exponential in vector lengths shorter than approximately 3mm.
Neither the wedge angle nor the offline and online tuned controllers are observed to affect the error trend. 

\begin{figure}[ht!]
\begin{center}
\includegraphics[width=1\columnwidth]{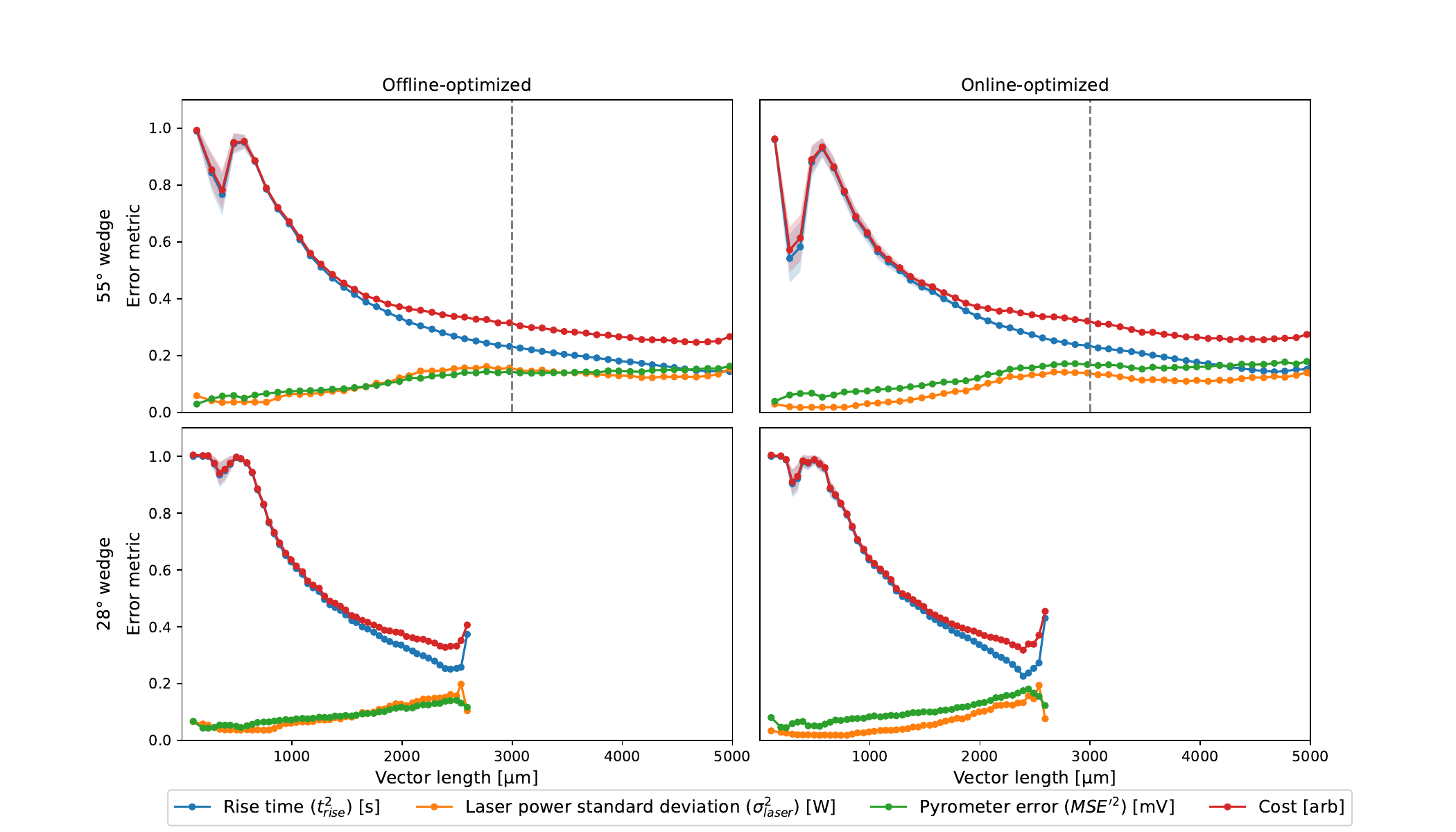}
\caption{Overall cost and each cost function term plot for varying vector lengths for all printed wedge parts. Each point is a single vector with its value calculated by averaging through all printed layers.}
\label{fig:costs_subfactors_comparison} 
\end{center}
\end{figure}

To further elucidate the individual contributions of each element of the cost function plotted in \Cref{fig:costs_comparison},  \Cref{fig:costs_subfactors_comparison} shows each metric separately for both geometries and both optimization methods.

Laser power standard deviation and the pyrometer error exhibit a constant trend in vectors larger than 3mm, as the rise time linearly increases.
The increase in rise time with constant other metrics yields a minor change in the overall cost value.
Shorter than 3mm vectors show a distinct change in the error components.
Pyrometer error and laser power standard deviation linearly decrease as the rise time starts to exponentially increase.
The overall cost value is observed to be dominantly defined by the rise time components in shorter vectors.

\section{Discussion}
\label{sec:discussion}

\subsection{Optimization progress}
With either the online- or offline-tuning procedure optimization settings, BO achieves the lowest cost value around the similar magnitude $P$ value with $\sim30\%$ deviation of the $I$ value.
As shown in \Cref{fig:optimization_iterations}, the algorithm achieved a lower cost value sooner in the online case than offline.
Faster error minimization in the online case can be attributed to the higher reference value assigned, which forces a higher $I$ value to obtain any error value smaller than one due to the normalization of the rise time.
Furthermore, the presence of powder increases the strength of the relationship between laser power and the pyrometer signal relative to a bare plate.
Naturally, the same laser power input yields a higher emission value due to the lower mass density per applied energy in the powder case. 
The effect of the powder and the steel plate on the lowest achieved error during optimization requires further study to be fully evaluated.

For the demonstration of the proposed method, the number of iterations was conservatively fixed to 200.
The optimization progress is assumed to have minimized the cost function based on non-decreasing error values in the later iterations.
However, there several approaches that can be implemented as a stopping strategy for decreasing the total tuning duration~\cite{ishibashi2023stopping}.
Overall, the offline tuning takes approximately 3 seconds per iteration and took 10 minutes to perform 200 iterations.
The online tuning method heavily depends on the machine and the build setup for the optimization application, however, online tuning adds an insignificant amount of extra time to the normal build cycle.
Choosing an online or offline optimization strategy for controller tuning is a decision that mostly depends on the availability of hardware since the controllers tuned by both procedures achieve similar performance.
On the contrary, manual tuning of the setup by already proven heuristic methods takes several hours with a domain-expert operator.
Therefore, the proposed algorithmic tuning represents a significant step forward for the state-of-the-art controller tuning applications for the LPBF process.

\subsection{Optimization-signal characteristics}
As shown in \Cref{fig:best_iterations_combined}, there are several noteable differences in the pyrometer signal between the optimization with and without powder.
The differences such as downward and upward trends at the end of the vectors in offline and online cases can also be explained by the substrate anomalies.
The substrate profile may have distorted due to overexposure of the same region in offline and the build-up of a single-vector line in online settings.
These anomalies primarily affect the mean squared error term of the cost function, however, they are observed to be consistent across the layers, and as such do not significantly affect the optimization process.
A significant finding is the overshooting phenomenon that only occurs in the powder exposure case which is shown by the magnified region in the online control case as shown in the bottom row of \Cref{fig:best_iterations_combined}.
The signal overshoots the reference value by $7\%$ and continues to fluctuate until it settles to the reference value over the first third of the vector.
Although the signal remains within $\pm 5\%$ of the reference value, the fluctuation is observed to continue until settling $\sim$3mm (around 6ms) from the start of the vector.
The same overshooting and settling behavior is observed in the laser power assignment and the pyrometer absolute error.
As described in \Cref{sec:cost_function} and in \eqref{eq:mse}, the mean squared error of the pyrometer signal is calculated only for the second half of the signal to assess the tracking accuracy without the influence of the melt pool initiation transient.
Therefore, the cost function is blind to the fluctuation in the first half of the vector, and optimization progress is not affected by the fluctuation.
Aside from the overshooting and settling differences, the rise times for the offline- and online-optimized cases are 920 and 1140 microseconds, respectively, which can be attributed to the difference in the $I$ values in both optimal cases 

\subsection{Wedge prints: PI control implications}
The time series data from the wedge prints in \Cref{fig:wedge_28_timeseries} and \Cref{fig:wedge_55_timeseries} show the vector-to-vector trend of the controller, namely decreasing the laser power as the vector size decreases to track the pyrometer reference value.
The overshooting and delayed settling phenomenon observed in online optimization is still evident at the in-vector scale.
The magnified image of the pyrometer signal of the offline case of 55° wedge geometry in \Cref{fig:wedge_55_timeseries} (top left) clearly shows the signal overshoot before it settles by the end of the first half of the vector.
Once the vectors are shorter than $\sim$3mm (i.e., approximately halfway through the time series representation), the pyrometer signal no longer settles fully. 
The non-settling behavior of the signal represents an oscillatory behavior in a vector-to-vector case, as shown by the magnified pyrometer signal of the online-optimized case in \Cref{fig:wedge_55_timeseries} (top right).
The same behavior is observed through the entire layer of the 28° wedge, as shown in \ref{fig:wedge_28_timeseries} due to the shorter initial vectors ($<$3mm). 
The spatial distribution of the overshooting and short vector oscillatory behavior is also observed in the scatter plots shown in \Cref{fig:scatter_50} and \ref{fig:scatter_28}, and is especially made evident by the color bands of the lower plots. 
For vectors of the 55° wedge, the pyrometer value starts too low, but then it quickly overshoots before stabilizing to the set point.
The online-tuned controller appears to overshoot and undershoot more than the one tuned offline. 
The length of the initial transient appears to be constant along the progression of the hatches, despite the decreasing vector length.
The spatial distribution of the pyrometer signal also aligns with the overshooting and shorter vector oscillation observation in the time scale.
Overall, the findings suggest a minimum controlled vector length of approximately 3mm using the experimental settings of this study.
However, assuming that the pyrometer signal can be differentiated without introducing excessive noise, the overshoot may be further decreased by including a derivative control term to create a PID.
The improvement of the signal quality, whether by hardware improvements or digital filtering techniques, is outside of the scope of this study.

\subsection{Wedge prints: controller-induced porosity}

In \Cref{fig:microstructure}, one of the wedge part's microstructural evaluations is shared.
Shared findings are found to be representative enough for the rest of the parts.
It is observed that the vectors starting at the right side (marked with B) show no porosity formations.
This confirms that the resulting laser power assignment to track the selected reference value is within the nominal process window.
However, overheating introduced by shorter vectors (towards point A) increases the melt pool emissions so that the laser power is driven lower by the controller to maintain the emission intensity at the reference value.
Lowered laser power results in insufficient energy, therefore, the process shifts out of the process window to the lack-of-fusion zone.
Unconstraint energy input with an overheating process condition is expected to violate the process window due to the narrow process window of the high-density microstructure compared to the wide range of the heat accumulation phenomenon as expressed by the uncontrolled part shown in \Cref{fig:scatter_28}.
Conclusively, the results suggest that an in-layer dynamic laser power control with a static reference value is not generalizable for all geometries.
Expanding the controllable limits by maintaining the energy input within the process window represents a challenge and a promising future research direction for the implementation of control strategies.

\subsection{Wedge prints: cost function and error analysis}
All wedge prints exhibit similar and consistent overshooting and delayed settling behavior for both the online- and offline-tuning cases, regardless of the vector size. 
A more in-depth comparison of the effect of the vector length on each of the cost function terms along with the final cost value for each wedge print case is shown in \Cref{fig:costs_subfactors_comparison}.
For both the online- and offline-optimized parameter settings, the 55° wedge error metrics are relatively constant until the vector length decreases below approximately 3mm.
Shorter vector lengths drive the laser power and pyrometer error values lower and increase the rise time cost value. 
The overall cost value is dominantly defined by the rise time in all vectors shorter than 3mm.
An important remark is that as described in Section \ref{sec:cost_function}, all cost terms are normalized by the vector length. 
While the pyrometer signal's MSE and laser's standard deviation components describe the tracking performance of the controller through the vector, the rise time is defined only at the beginning of each vector and is independent of the size of the vector in a part printing.
Due to the independence, shorter vectors yield significantly higher rise time components and ultimately cost terms.
Despite the bias introduced by the normalization of the rise time component, the same rise time component is used for the consistent comparison of the online and offline cases as well as to evaluate the representativity of the optimization signal used in the BO algorithm of the real part printing.
The constant normalization factor $C$ added to the error terms functions as a weight term.
It normalizes the different units and magnitudes of error terms for a similar representation in the cost calculation.
Hence, $C$ is expected to change the effect of each error subterm to the overall cost value calculated in each iteration.
The term is proposed to be defined concerning the reference value and the laser exposure strategy used in the experimental setup, therefore, should be adjusted depending on the implementation.

\subsection{Online- vs offline-tuning}
No significant difference between the wedges of online- and offline-optimized parameters is observed, which can be attributed to the relative similarity of the optimal parameters defined for either method as shown in \Cref{fig:costs_comparison}.
The difference in the integral gain does not change the signal characteristics in either of the wedge prints apart from a marginal difference in the laser power standard deviation component of the cost function of the online optimized wedges as shown in \Cref{fig:costs_subfactors_comparison}.
A lower laser power standard deviation term can be attributed to the effect of higher integral gain on the stability of the controller.

As suggested by the similarity of the findings of online and offline experiments, algorithmic tuning of a high-frequency laser power controller can be performed using either method with similar results.
The optimization procedure can be applied autonomously at the beginning of a build within the range of the sacrificial layers (i.e., due to build platform separation) before initiating the controller with the optimized parameters after the defined number of layers as iterations. 
No prestudy or material-specific knowledge is required besides the initial gain assignment for the optimization process since the method proposed in this study is a black-box optimization algorithm that minimizes a cost metric by screening the defined inputs.
Similarly, tuning performed on the artificial setup is also transferable to the controller for the actual process.
The offline optimization plate is also a steel alloy S304 unlike the powder used 316L.
Although the material is not the same, using a different alloy system may not be representative of the signal characteristics.
Each approach could be employed depending on the availability of the dedicated hardware for specific applications.

\subsection{Parameter window and reference value selection}

The selected wedge geometries exhibit heat accumulation due to the decreasing vector-to-vector exposure times with the decreasing vector lengths.
The geometries are specifically selected to represent a commonly occurring in-layer heat accumulation phenomenon in printed geometrical features.
Increasing preheat increases the melt pool emissions and the controller decreases the laser power input as a result.
Due to the changing laser power input, two factors are reported as critical to the quality of the printed parts: Reference value selection and the parameter window.
The reference value is required to capture the nominal process behavior, i.e. the emission observed under non-overheating conditions.
In \Cref{fig:wedge_55_timeseries}, it is observed that while compensating for the overheating, the laser power is driven below the parameter window at 140W as described by the dashed line, which results in lack of fusion porosities in the microstructure, as expected.
The study of Kavas et al. described a limit for controlling the temperature on a layer-to-layer scale by actuating the laser power value due to the process window~\cite{kavas2023layer}.
The limit of a similar root cause is observed in this study on the vector-to-vector scale.
The vector-to-vector heat build-up behavior is stabilized with the expense of transiting the lower boundary of the process window.
This observation highlights the necessity of a cooling-based controller design where the cooling of the printed part through waiting is also actuated as a promising future study on an in-layer scale.
Similarly, a higher reference would enable the controller to avoid descending to the lack-of-fusion zone.

\subsection{Future work on in-layer control}

The experimental findings of the proposed method in this study suggest multiple directions for future studies to further enable closed-loop control applications for the LPBF process.
First, for a more robust in-layer controller tuning application, the approach maybe enhanced with the addition of the derivative (D) term, and the cost function can be improved by including further signal quality metrics such as overshoot and settling time.
This assumes that the signal is of sufficient quality that the noise generated by discrete differentiation does not detract from the performance already achieved with a PI controller.
A new parameter that adds a variable vector-to-vector dwell time for reducing heat accumulation and maintaining the input within the process window could further improve the performance.
Furthermore, despite the hardware and computational complexity challenges, other available monitoring sensors such as high-speed or thermal cameras can be implemented into the closed-loop control application to stabilize the melt pool for various metrics such as size and shape.
Similarly, the correlation of the pyrometer measurements with melt pool geometries and solidification rates could be used to tune the controllers for specific microstructure. 
The BO algorithm used in this study can further be applied for sensor-based parameter tuning of novel alloys in a closed-loop manner to reduce the process parameter optimization costs.

\section{Conclusion}
\label{sec:conclusion}
A Bayesian Optimization algorithm for was implemented for tuning an in-layer closed-loop controller for the the LPBF process and experimentally demonstrated using PI control of the laser power for meltpool stabilization.
Furthermore, online- and offline- optimization approaches were proposed to enable fast and efficient optimization and were experimentally validated.
The proposed cost function components are the mean square root error of the pyrometer signal, the pyrometer signal's rise time, and the variance of the laser power assignment around the rolling average.
Offline optimization yielded a lower error in 112 iterations while the online optimization converged earlier at iteration 74 in the experiment.
The tuned controllers were experimentally employed to control the heat-accumulating behavior of two wedge geometries with different angles to evaluate the controller performance on in-layer heat accumulation settings.
The results showed similar performance on both parts and by offline and online controller settings while efficiently controlling the in-layer temperature compared to the non-controlled example.
The controller's ability to stabilize the pyrometer signal was observed to be correlated with the vector length and the signal error increases with decreasing vector length. 
the minimum controllable vector length is determined by the transient state of the process in which the rise time determines the cost function; for the material studied, this yields a minimal controllable vector length of 3mm.
Moreover, the findings of this study highlight the importance of appropriate reference value assignment and the process window restriction.
Compensating the overheating in wedge geometries drove the laser power outside of the process window, which introduced lack-of-fusion defects while stabilizing the melt pool emissions to the given reference values.
Future work is proposed to correlate the pyrometer signal with the microstructure of the part such that the controller can be tuned for different objectives.
Implementation of varying reference values for different regions of the part to maintain the laser power inside the parameter window with the expense of allowing overheating can also increase the robustness and applicability of emission-based closed-loop control applications.



\bibliographystyle{model1-num-names}

\bibliography{cas-refs}

\end{document}